\documentclass[10pt]{article}
\usepackage{simpleConference}
\usepackage{times}
\usepackage{graphicx}
\usepackage{amssymb}
\usepackage{url}%

\usepackage{hyperref}

\usepackage{amsmath}
\usepackage[numbers]{natbib}
\usepackage[utf8]{inputenc}
\usepackage[T1]{fontenc}
\usepackage[british]{babel}
\usepackage{enumerate}
\usepackage{geometry}
\usepackage[margin=1cm]{caption}
\usepackage{paralist}
\usepackage{fancyhdr}
\usepackage{here}

\usepackage{siunitx}
\usepackage{graphicx}

\DeclareMathOperator*{\argmin}{arg\,min}

\begin{document}

\title{Optimal allocation of defibrillator drones in mountainous regions
\thanks{
This work was supported by the European Union Fond for regional development and Interreg V-A Italy Austria 2014-2020 and was carried out while the first author was employed within the Interreg project START (Smart test of Alpine rescue technology).
}
}

\author{
C. Wankm\"uller
\thanks{Department of Operations Management and Logistics,
Alpen-Adria-Universit\"at Klagenfurt,
Austria,
\href{mailto:christian.wankmueller@aau.at}{christian.wankmueller@aau.at} }
,
C.\
Truden\thanks{Department of
Mathematics, Alpen-Adria Universität Klagenfurt, Austria,
\href{mailto:christian.truden@aau.at}{christian.truden@aau.at}}
,
C.\ Korzen
\thanks{Department of Operations Management and Logistics,
Alpen-Adria-Universit\"at Klagenfurt,
Austria,
\href{mailto:christopher.korzen@aau.at}{christopher.korzen@aau.at} }
,
P.\ Hungerländer\thanks{Department of
Mathematics, Alpen-Adria Universität Klagenfurt, Austria,
\href{mailto:philipp.hungerlaender@aau.at}{philipp.hungerlaender@aau.at}}
,
G.\ Reiner
\thanks{
Institute for Production Management, Vienna University of Economics and Business, Austria,
\href{mailto:gerald.reiner@wu.ac.at}{gerald.reiner@wu.ac.at}
}
, and
E. Kolesnik
\thanks{ Division of Cardiology, Medical University of Graz, Austria,
\href{mailto:ewald.kolesnik@medunigraz.at}{ewald.kolesnik@medunigraz.at}
}
}

\maketitle
\thispagestyle{empty}

\begin{abstract}
  Responding to emergencies in Alpine terrain is quite challenging as air ambulances
  and mountain rescue services are often confronted with logistics challenges and
  adverse weather conditions that extend the response times required to provide life-saving
  support. Among other medical emergencies, sudden cardiac
  arrest (SCA) is the most time-sensitive event that requires the quick provision
  of medical treatment including cardiopulmonary resuscitation and electric
  shocks by automated external defibrillators (AED). An emerging technology called
  unmanned aerial vehicles (or drones) is regarded to support mountain rescuers in overcoming the time criticality of
  these emergencies by reducing the time span between SCA and early defibrillation.
  A drone that is equipped with a portable AED can fly from a base station to the
  patient’s site where a bystander receives it and starts treatment.
  This paper considers such a response system and proposes an
  integer linear program  to determine the optimal allocation
  of drone base stations in a given geographical region. In detail, the
  developed model follows the objectives to minimize the number of used drones
  and to minimize the average travel times of defibrillator drones responding to SCA patients. In an
  example of application, under consideration of historical helicopter response times, the authors test the developed
  model and demonstrate the capability of drones to speed up the delivery of
  AEDs to SCA patients. Results indicate that time spans between SCA and
  early defibrillation can be reduced by the optimal allocation
  of drone base stations in a given geographical region, thus increasing
  the survival rate of SCA patients.
\end{abstract}

\setlength{\tabcolsep}{6pt}
\renewcommand{\arraystretch}{1}

\section{Introduction}\label{sec:intro}
In recent years, tourism in the Alps has gained increased popularity
through easier access to previously untouched regions for mountaineers. Simultaneously,
the number of medical incidents in mountainous regions has increased sharply
\cite{bergrettung}. Under these circumstances, emergency response is quite challenging, as response teams are often confronted with severe logistics challenges and weather
conditions that cause long response times for life-saving support. \par
Among the different sets of medical emergencies
(e.g. accidents while mountain biking, climbing, rafting etc.),
sudden cardiac arrest (SCA) accounts for a remarkable proportion of lethal events.
For instance, in the last ten years $707$ people died in the Austrian Alps, with $37\, \%$ due to SCA
representing the highest proportion of all lethal incidents \cite{Alpinunfall}.
The immediate stop of cardiac activity during SCA leads to a rapid collapse
of all vital organ processes and further, inevitably, death if untreated.
The European Resuscitation Council (ERC) provides evidence-based guidelines
for the best available treatment of an SCA
event, namely, early recognition
and call for help, immediate cardiopulmonary resuscitation (CPR), as early as
possible electrical defibrillation with an automated external defibrillator (AED),
professional advanced life support, and standardized post-resuscitation care \cite{monsieurs2015european}.
The latter two steps in this chain of survival require well-educated
emergency medical service (EMS) and their effectivity is largely dependent
on the previous steps, because the EMS measures will inevitably arrive after a certain time delay.
However, immediate response to an SCA event can only be provided by bystanders.
According to the currently valid formula from \cite{larsen1993predicting},
survival rates decline at a rate of approximately $8\text{-}16 \, \%$ per minute without CPR
and electrical defibrillation. If CPR and electrical defibrillation
are performed properly, survival rates decline less dramatically at
approximately $3\text{-}4 \, \%$ per minute. In this context, bystander-use of an
AED is associated with better survival and functional outcomes \cite{pollack2018impact}.
Especially the rapid use of an AED - even without CPR - seems to be
beneficial and can enhance the life expectancy of SCA patients more than
2-fold \cite{Capucci2016}. According to \cite{nichol1999cumulative}, early
defibrillation should be performed within a time interval of $6$ minutes as
survival rates up to this threshold are nearly constant and sharply decrease
if no shock is provided within $6$ minutes.
In urban areas, programs that have been established to provide bystanders quick access
to AEDs in public places have led to an increased number of survivors after SCA in
public locations \cite{public2004public,weisfeldt2010survival}.
Moreover, a dense EMS network guarantees best available life support with
little time delay. In urban regions, recommendations set the time
between EMS notification and EMS arrival to a maximum of $20$ minute,s which
is almost impossible to follow  in rural areas \cite{bos2015ambulance}.
However, responding to a patient with SCA in Alpine regions is often extremely
challenging due to rough terrain.
Difficult access to the patient's site may lead to long
response times for mountain rescue services and even ambulance helicopters,
negatively affecting the chance to survive SCA events. Obviously, a reduction
of the time between the SCA event and proper treatment including AED shock therapy and CPR is the only way to increase the survival rate of patients.\par
An emerging technology called unmanned aerial vehicles (UAV), or drones,
is regarded to overcome the time criticality to respond
to such emergencies in mountainous regions.
Drones are flying machines that operate autonomously or are teleoperated by
ground operators \cite{floreano2015science}. The main drone types available
on the market are fixed-wing systems (comparable to a miniature plane), multirotor systems, and hybrid systems with different specifications
(i.e. payload, drop-off system, range) \cite{custers2016future}.
Conventional drones run on batteries to operate their engines and need to be
recharged after they run out of power. Therefore, base stations are used that allow
autonomous wireless recharging without the need of active user intervention
\cite{choi2016automatic}.
Aside from military applications, drones are attracting increasing attention in commercial
usage, including cargo delivery, mapping, target covering, or surveillance
\cite{dorling2017vehicle,di2015energy,murray2015flying}.
From a logistical viewpoint, they offer the clear advantage of being able to travel to areas
that are inaccessible for land-based transportation at lower cost and risk
compared to traditional means of transport (e.g. helicopters) \cite{kharb2015moving,yildirim2new}.
Recently, humanitarian organizations have also become aware of these advantages,
as drones can contribute to more efficient emergency operations along the
entire disaster management cycle, i.e. mitigation, preparedness, response,
and recovery stages \cite{Kakaes,anbarouglu2019drones}. Drones can assist in damage assessment,
emergency items delivery, and search and rescue  missions during the
immediate response to earthquakes or avalanche events
\cite{cui2015drones,doherty2007uav,mersheeva2012routing,camara2014cavalry}. \par
\begin{figure} [!ht]
  \begin{center}
    \includegraphics[width=0.45\textwidth]{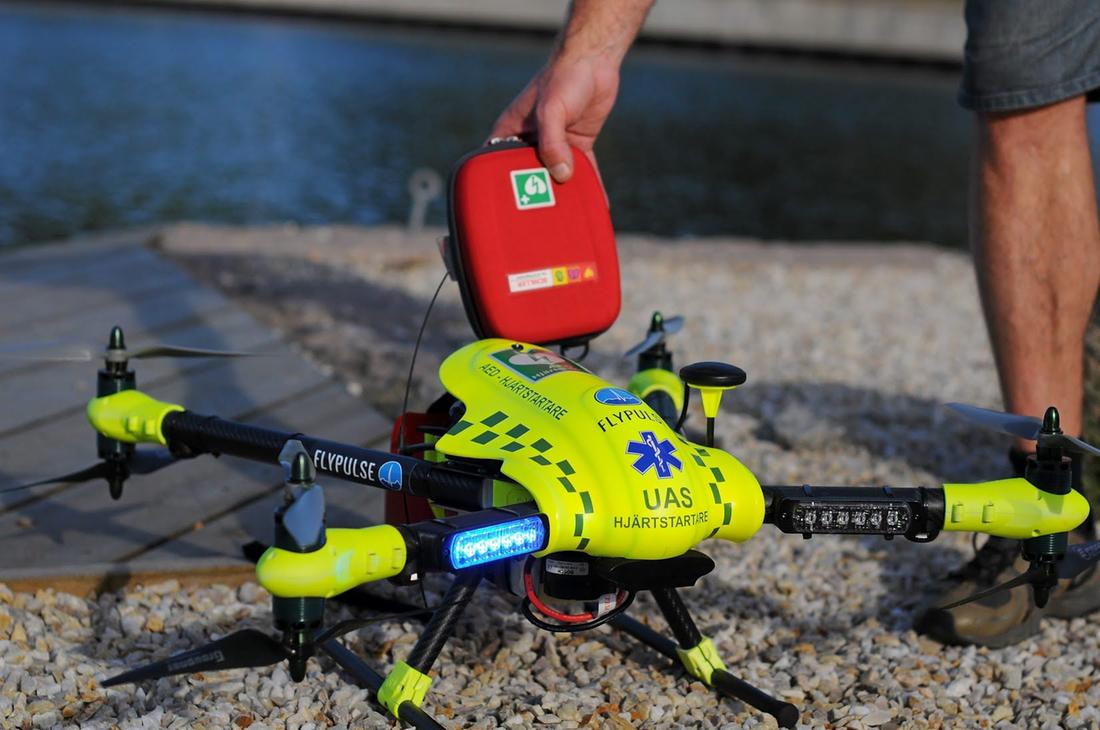}
    \caption{AED attached to the drone (\textit{LifeDrone AED}). }
    \label{fig:drones}
  \end{center}
\end{figure}

Especially, the drones' capability to transport various kinds of relief items to
demand locations is of major interest in this study. Recent developments in drone
technology enable not only the delivery of lightweight items (i.e. vaccines,
blood samples, etc.) but even AEDs of heavier weight. In this regard, the drone
either delivers an attached portable AED (e.g. \textit{LifeDrone AED}, see Figure \ref{fig:drones}) or both modules are
merged within one technical unit (e.g. \textit{Ambulance Drone} developed by TU Delft \cite{ambulanceDrone}).
In practice, AED drones are activated once an alert arrives at the emergency
coordination center. The required GPS (global positioning system) coordinates
are either actively transmitted by the bystander via a mobile phone app to the emergency coordination center or generated by tracking the mobile phone of the caller. The GPS data is then sent to the drone base station and processed by the defibrillator drone which departs to the patient's site.
Once the bystander receives the defibrillator drone, he/she disconnects the AED from it and starts adequate treatment.
The possibility to disconnect the AED from the drone ensures that it can be transported even to forest
areas where trees eventually hamper safe drone landings. With video transmission
between the drone and emergency operators, the bystander is guided through
the whole process of providing CPR and putting electrodes on the patient's chest.
This response system could serve as an extension of ground-based (i.e. mountain rescue service)
and air ambulance. It offers the clear advantage of providing faster emergency help if needed.
Hence, it has the potential to reduce the time between SCA and early defibrillation, which is essential for
increasing patients' survival rates, outcomes, and quality of life \cite{zegre2018delivery}.
A well-planned network of drone base stations in a given geographical
region can guarantee minimal response times (referred to as travel times in this paper) to SCA patients.
Especially, in large geographical
areas for which a single drone application would not generate beneficial
outcome for the patients, an optimally allocated network of defibrillator drones is required. However, the battery capacity, which limits the range of the drones, must be considered.
Motivated by exploring the potentials of drone technology to speed up live-saving support, this paper considers drones as a means to deliver
AEDs in Alpine terrain and presents an integer linear program for the optimal allocation
of drone base stations to minimize the travel time of defibrillator drones for providing
life-saving support in the minutes after SCA.
\par The paper is organized as follows. First, we review related
papers that present models for the optimal allocation of drones for the delivery
of AEDs to SCA patients. Next, we give a problem description and introduce notations
along with our model assumptions. Then, we present our optimization model and
illustrate its use in an example of application. A discussion on the limitations of the study and an outlook to future research conclude the paper.

\section{Related Work}
The presented research is built upon existing works on location allocation problems
in humanitarian logistics, which have been extensively researched over the past years.
A broad range of different problems, including the optimal allocation of fire
stations, EMS services, or medical centers were solved using heuristic and
exact algorithms. These models follow the overall objectives to minimize the
total distance between the demand points and candidate facility locations, to maximize
the total number of demand points covered within the distance limitations, or
to minimize the worst system performance
\cite{boonmee2017facility}.
With regard to the optimal allocation
of base stations of aerial vehicles (e.g. helicopters and drones),
models have been proposed by \cite{talmar2002location, roislien2017exploring, schuurman2009modelling}.
Reference \cite{talmar2002location} focuses on optimizing the location of three rescue
helicopters in South Tyrol to be able to respond to the increasing number of skiing, hiking
and climbing accidents in the region.
The objectives of the developed model
are to minimize the
maximum or worst response time to speed up emergency operations.
Another model for improved helicopter response times is presented
by \cite{schuurman2009modelling}. The authors propose a location
optimization model to identify where the expansion of helicopter EMS
covers the greatest population in British Columbia among those people
that were under-supplied during that time period.
In \cite{roislien2017exploring} a model to estimate optimal locations for air
ambulance service in Norway is proposed.
The developed maximum coverage location problem (MCLP) model maximizes the
population covered within a specific service distance (or time) by optimally
allocating a pre-defined number of facilities. Optimal allocation of base
stations that operate drones and trucks are proposed by \cite{chowdhury2017drones}.
The proposed model determines the optimal distribution location for drones and trucks to
deliver emergency items to a set of demand locations within a disaster affected area.
Therefore, the model minimizes the total location-routing cost to serve the
whole region by drones and trucks.
\par
Models that are more relevant to our study and exclusively focus on the optimization of the
spatial location of defibrillator drones are presented in \cite{pulver2016locating, boutilier2017optimizing, pulver2018optimizing}.
Reference \cite{boutilier2017optimizing}
develops an integer optimization model to determine the number of
drone base stations for reducing the historical median fire, paramedic, and police response
times by $1$, $2$, and $3$ minutes. Aside from identifying the optimal location
of drone bases stations, the authors also seek to determine the optimal
number of drones for each station. Historical data including more than 53,000 SCA events
that occurred between 2006 and 2014 in the eight regions of Toronto Regional RescuNET was
used to test the developed optimization model. All fire, paramedic, and police stations of the
region were chosen as candidate locations for defibrillator drones. The results indicate that optimized
drone networks can considerably reduce response times to provide early defibrillation.
In \cite{pulver2016locating} a network of defibrillator drones designed to minimize the response time to SCA in Salt Lake County is presented.
The objective is to have a drone at the scene within one minute for at least $90 \, \%$ of SCA
incidents while also minimizing implementation costs.
The developed MCLP model determines the optimal configuration of the drone network
by maximizing the total demand suitably covered.
The results indicate that current EMS is only capable of reaching the scene
within one minute in $4.3 \, \%$ of all SCA incidents whereas including drones in the
response system leads to $80.1 \, \%$ of demand being reached within one
minute.
Further, installing additional launch sites would lead to $90.3\, \%$ covered
demand within one minute. Reference \cite{pulver2018optimizing} is based on the model
proposed in \cite{pulver2016locating} and presents  an extended model considering backup service provision,
continuously distributed demand, and
empirical medical data instead of estimated incidence rates.
Based on these extensions the authors propose a new spatial optimization model,
namely the backup coverage location problem with complementary coverage.
The model objective is to maximize the total amount of primary and backup
coverage for demand, i.e. SCA events.
The model is applied in the same setting
as in \cite{pulver2016locating} and generates more accurate results by
mitigating representation errors in locating a network of defibrillator drones.
\par
To the best of our knowledge, no other work could be identified that
addresses the optimal allocation of base stations for defibrillator drones in
equal measure as presented in this paper.
Our study is unique in the sense that aforementioned papers treat defibrillator drone
allocation in a purely urban environment, while we introduce this concept to
mountainous regions for the first time. This setting is different compared to
already discussed ones due to infrastructural and geographical barriers
that need to be considered in the determination of the optimal allocation
of drone base stations. In terms of base station selection, our work differs to
others in the sense that we cannot use EMS stations as candidate locations due to
their low availability in mountainous regions. Instead, we integrate Alpine shelters
and fire rescue stations as potential drone base stations.
Alpine shelters are directly available within Alpine infrastructure and
the only permanently installed housing structure in such rural areas.
Voluntary fire rescue stations are generally much more available in Alpine infrastructure than EMS stations, as they are also maintained in really remote areas.
If an SCA patient is located close to a village, a defibrillator drone departs
from a close-by fire rescue station instead of an Alpine shelter hut.
Travel times from a high-altitude base station would be extremely long,
resulting from the differences in altitudes.
Compared to other models, we primarily ignore backup supply and argue that
two SCAs in a rural area at the same time are rather unlikely to occur.
However, as this case cannot be completely excluded, we also extend our model such that backup supply can be considered.
Finally, we compare our approach against the conventional air ambulance system in an example of application. Other papers follow a comparison with
ground-based EMS that would not fit the context of this paper.

\section{Optimization Approach} \label{sec:prob}

We now introduce the integer linear program (ILP) that we use to model
and analyze the allocation of drones to base stations.

Due to the characteristics of mountainous regions the  travel times of aerial vehicles, especially drones, can differ significantly
even for sites that are close in terms of latitude and longitude.
Peaks and cliffs form time-consuming obstacles that the drones must overcome.
Hence, the coverage of a considered region is modeled by sampling
a large number of patients and  calculating the point-wise distances
between all patients' sites and all possible drone base stations.
The considered problem is a paradigm for the well-known
\textit{simple plant location problem} (SPLP) that is also referred to as
\textit{uncapacitated facility location problem} (UFLP) \cite{KLOSE20054, locationScience}.
Considering the analysis presented in this work, solving an ILP (introduced in Subsection \ref{ssec:ilp})
is the method of choice for the occurring problem size.
For even larger problem instances we refer to heuristic approaches from the literature
\cite{BARAHONA200535, GOLDENGORIN2003967, Joernsten2016,LETCHFORD2014674, Vecihi2006}.

\subsection{Model Parameters \& Assumptions}

\paragraph{Basic Parameters.}
Formally, the problem is defined through the following parameters:
\begin{itemize}
  \item A set of $m$ \textit{candidate locations} (base stations)  $\mathcal{B}:=\{b_1,\ldots,b_m\}$.
  The coordinates of a candidate location $b \in \mathcal{B}$ are given by $(x_b,y_b,z_b)$
  correspond
  to latitude, longitude, and altitude.
  \item A set of $q$ \textit{patient's sites}  $\mathcal{P}:=\{p_1,\ldots,p_q\}$.
  The coordinates of a patient $p \in \mathcal{P}$ are $(x_p,y_p,z_p)$.
  \item A \textit{travel time function} $t: \mathcal{B}  \times \mathcal{P} \mapsto \mathbb{R}^{+}$.
  \item A \textit{homogeneous fleet}  of $s$  \textit{drones}   that must be
  assigned to the base stations $\mathcal{B}$. Clearly, $s \leq m$.
  \item
  A maximal travel time $t_{max}$ that cannot be exceeded.
\end{itemize}

In Table \ref{tab:parameters} we summarize the used notation.

\begin{table}[!ht]
  \begin{center}
    \begin{tabular}{ l| l | l }
      & Definition & Description \\
      \hline \hline
      $\mathcal{B}$ &  $\mathcal{B}:=\{b_1,\ldots,b_m\}$ & candidate locations \\
      $\mathcal{P}$ &  $\mathcal{P}:=\{p_1,\ldots,p_q\}$ & patients' sites \\
      $t$ & $t: \mathcal{B}  \times \mathcal{P} \mapsto \mathbb{R}^{+}$  & travel time function \\
      $s$ & & number of available drones \\
      $t_{max}$ & & maximal allowed travel time
      \\
      \hline
    \end{tabular}
    \caption{List of input parameters and constants.}
    \label{tab:parameters}
  \end{center}
\end{table}

\paragraph{Model Assumptions.}
In order to allow mathematical modeling of the problem,
the following assumptions are made:
\begin{compactitem}
  \item No more than one SCA happens at any one time in the same region, therefore no backup is needed.
  \item Weather and wind conditions are good enough for the selected drone models
  to operate. The average wind speed in the Alps is around
  $9.5\,m/s$ \cite{windanalysis}. Both drone models considered in the subsequent study can operate under these wind conditions (see Table \ref{tab:droneParameters}). In fact, drones (and also helicopters) are not always
  suitable to transport AEDs as wind speeds up to $34.2 \, m/s$
  are possible under stormy conditions.
  Therefore, alternative transport solutions must be considered
  (e.g. AEDs are part of mountain rescue service equipment).
  \item All patients $p \in \mathcal{P}$ are localized
  on an official hiking trail. We assume this because hiking trails are the major
  routes that are followed by hikers, thus the likelihood of having SCA patients in
  such areas is greater compared to remote areas far away from the official hiking trail network.
  In comparison, more than $80 \, \%$ of all fall-related accidents happen on marked
  trails \cite{faulhaber_fall-related_2017}.
  \item The maximal reach (battery capacity)
  of the considered drone is larger than the
  maximal distance that it can travel within the maximal travel time $t_{max}$.
\end{compactitem}
We would like to highlight the fact that these assumptions are empirically based. This supports the primary objective of the study to show that such a defibrillator drone system is applicable in mountainous regions utilizing rare infrastructure.

\begin{figure}[!ht]
  \begin{center}
    \includegraphics[width=0.8\textwidth]{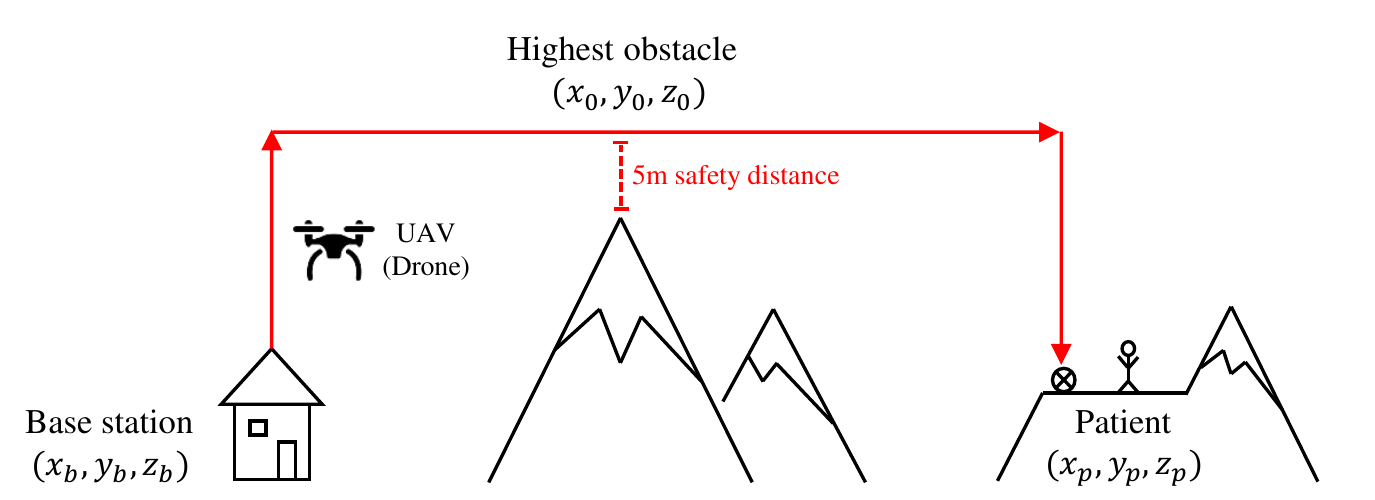}
    \caption{Illustration of the applied distance measure.}
    \label{fig:distance}
  \end{center}
\end{figure}

\paragraph{Travel Time Function.}
We calculate the travel time between a base station $b \in \mathcal{B}$ and a patient
$p \in \mathcal{P}$ using the following model.
Therefore, we consider the largest \textit{obstacle} in the direct line connecting $b$ and $p$, denoted by
$o(b,p)$   having coordinates $(x_o,y_o,z_o)$.
Further, we assume that the drone has a defined
\textit{vertical ascending  speed} $v_{vert}^{+}$,
a \textit{vertical descending  speed} $v_{vert}^{-}$,
and a \textit{horizontal travel speed} $v_{hor}$.
Moreover, the drone needs a constant \textit{start-up time} $c_{start}$, which defines the
time span from receiving the alarm until take-off.
The model assumes that the drone first rises vertically
to the altitude of the largest obstacle $z_o$ along its path
(plus $5 \,m$ safety distance).
Then it travels horizontally following the direct line between
the base station and the patient's site, which is described by the Euclidean distance between $(x_b,y_b)$ and $(x_p,y_p)$ \cite{hong2018range, ehrgott2002location, pulver2016locating}.
Finally, the drone descends vertically  to the patient's site. Hence, the travel time from $b$ to $p$ is given by:
$$
t(b,p):= c_{start}+
(z_{max}-z_b) \cdot v_{vert}^{+}
+ \sqrt{   (x_b- x_p)^2 + (y_b-y_p)^2} \cdot v_{hor}
+(z_{max}-z_p) \cdot v_{vert}^{-},
$$
where $z_{max}:=\max\{z_b,z_o+5 \,m ,z_p\}$.
In  Figure  \ref{fig:distance} we illustrate the model. Clearly, for practical application a more sophisticated way to determine the flight route could result in shorter travel times.

\subsection{Mathematical Model}
\label{ssec:ilp}
Using the notation and assumptions from the previous subsection we now
formulate the ILP.
The sets of indices $[u], u \in \mathbb{N}$, contain the elements $\{1,\ldots, u \}$.
First we introduce the \textit{assignment variables} $x_{ij} \in \{0,1\},~i\in [m], ~ j \in[q], $
with the following interpretation:
\begin{align*}
  x_{ij} &=
  \begin{cases} 1, & \text{if patient $p_j$ is assigned to base station $b_i$}, \\
    0, &    \text{otherwise}. \end{cases}
    \intertext{ Moreover, we introduce variables $y_i  \in \{0,1\}, ~ i \in [m],$ such that}
    y_{i} &=
    \begin{cases}
      1, &  \text{if a drone is located at base station $b_i$,}   \\
      0, &  \text{otherwise.}
    \end{cases}
  \end{align*}
  Consequently, we propose the following model:
  \begin{subequations}
    \label{form:fone}
    \begin{align}
      \min \quad & \alpha \cdot \sum_{i\in [m]} y_{i}   +
      (1-\alpha)\cdot \sum_{\substack{i \in [m],\\ j \in [q]} }   x_{ij} \cdot t(b_i,p_j)
      \label{eq:objective}
      \\
      \text{s.t.}\quad
      & \sum_{i\in [m]} x_{ij}  \geq 1,   \quad   j \in [q],
      \label{ineq:eachPatient} \\
      &\sum_{j\in [q]} x_{ij}   \leq  y_{i}\cdot q, \quad i \in [m],
      \label{ineq:onlyWhenDrone} \\
      & x_{ij} \cdot t(b_i,p_j) \leq  t_{max}  , \quad    i \in[m],~ j \in [q],
      \label{ineq:maxTime}\\
      & \sum_{i \in  [m]}  y_i  \leq s,
      \label{ineq:maxDrones}\\
      & x_{ij} \in \{ 0, 1 \} , \quad i \in[m],~ j \in [q],
      \label{set:patientAssignmentVars}  \\
      & y_{i} \in \{ 0, 1 \} , \quad i \in[q].
      \label{set:droneAssignmentVars}
    \end{align}
  \end{subequations}

  The objective function \eqref{eq:objective}
  allows to choose between two objectives by setting $\alpha$ to $1$ or $0$.
  Either the model minimizes the number of used drones
  or the average travel time.
  Inequalities \eqref{ineq:eachPatient} ensure that each patient is assigned to at least
  one base station.
  Inequalities \eqref{ineq:onlyWhenDrone} ensure that patients are only assigned to base
  stations that have a drone located there.
  Inequalities \eqref{ineq:maxTime} enforce that the defined time limit to reach a patient is not
  exceeded while Inequalities \eqref{ineq:maxDrones} guarantee that the available number of drones is
  not exceeded.
  Clearly, to ensure that the model has a feasible solution,
  each patient must be reachable from at least one base station in $t_{max}$ or less time,
  i.e.,
  \begin{equation}
    t_{max} \geq \overline{t}:=\max_{p \in \mathcal{P}} \min_{b \in \mathcal{B}}  t(b,p)
    \label{eq:minfeasibletime}
  \end{equation}
  must hold.

  \section{Example of Application and Results}  \label{sec:res}
\begin{figure}[!ht]
    \begin{center}
      \includegraphics[width=0.75\textwidth]{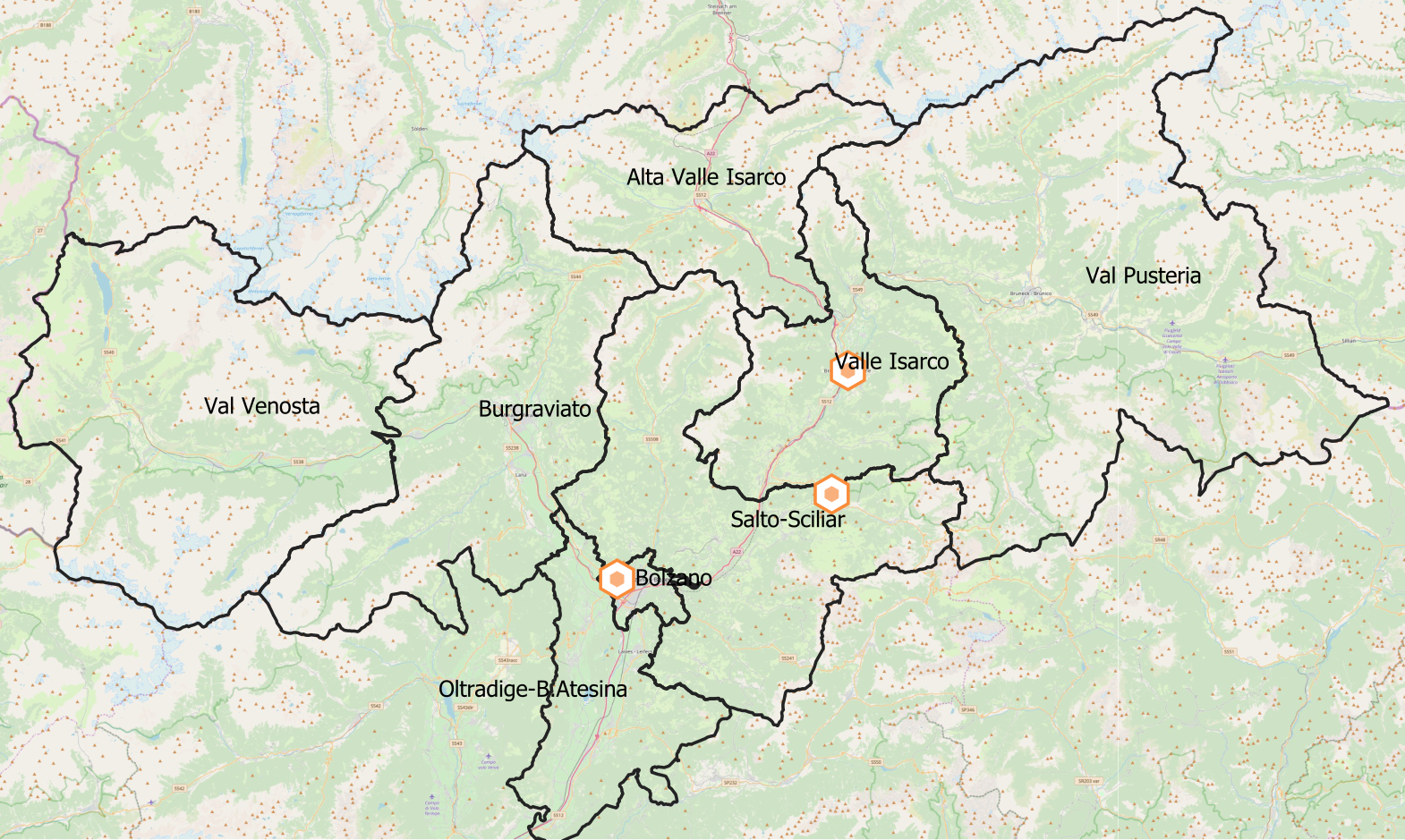}
    \end{center}
    \caption{Illustration of the seven districts of South Tyrol including the three helicopter base stations.}
    \label{fig:helis}
  \end{figure}

  In this section we examine the applicability of a network of base stations equipped with defibrillator drones in mountainous regions. First, we focus on \textit{Val Venosta}, which is
  the most western region of South Tyrol, Italy, covering an area of $1442\,km^2$.
  The region is crossed by many valleys and dominated by high mountains,
  including the Ortler, with its $3.905 \, m$ making it the highest mountain
  in South Tyrol. Obviously, this region represents the distinctive characteristics
  of mountainous areas, which are the major subjects of interest in this study.
  Furthermore, the region of South Tyrol is served by three ambulance helicopters
  with base stations in Bozen, Brixen, and Gröden that are geographically decoupled from the focus region Val Venosta (highlighted in Figure \ref{fig:helis}). This results in relatively long flight times for all three helicopters responding to patients located in Val Venosta. Installing a fleet of optimally located defibrillator drones in this region could therefore reduce the time between early defibrillation and helicopter arrival. We address this by analyzing the following two scenarios where in (1) we follow the objective to determine the optimal allocation of defibrillator drones in order to minimize the average travel times and in (2) we compare the travel times of defibrillator drones against conventional air ambulance in Val Venosta based on historical medical incidents in the region.

  \subsection{Data Preparation}
  \label{subsec:dataprep}
  Regarding the required data, we use local hiking trail network geographic information system (GIS) data
  provided by the state administration of South Tyrol \cite{wanderkartenSuedtirol}.
  For identifying available shelter huts in the region, we accessed three
  official data bases from the tourism office of South Tyrol and extracted
  the locations of available shelter huts \cite{vinschgau,AlmenTirol,Schnalstal}.
  In order to guarantee the validity of the locations of the identified shelter huts,
  we cross-checked them in Google Maps and OpenStreetMap and were able to locate
  additional shelter huts that were not included in the official databases.
  We derive data on fire rescue stations from the South Tyrolian
  association of voluntary fire brigades \cite{FF}.
  Finally, the corresponding latitudes and longitudes (World Geodetic System 84  - WGS 84) of all locations
  were gathered.
  Elevations for both types of locations as well as the elevation of the tallest
  obstacle in-between them were taken from the Google Maps Elevation API.
  Everything was implemented in Java and the ILPs were solved using Gurobi 8.1.0
  For illustration purposes, the GIS software QGis was used. For data analysis we used R.

  \subsection{Considered Drone Model}
  We operate the \textit{LifeDrone AED} system as described in \cite{claesson2017time},
  having parameters $v_{vert}^{+}=2.5 \, m/s$, $v_{vert}^{-}=2.5 \, m/s$, and
  $v_{hor}=17.9 \,m/s$ and $c_{start}=30 \,s$.
  The drone's own weight is $5.7 \, kg$ and it is capable of carrying an
  AED with a weight of $763 \,g$.
  The drone is equipped with GPS and
  autopiloted, hence it flies completely autonomously from the base station
  to the patient's location.
  It is notable that in the region of South Tyrol a smart phone app is already available that allows to transmit the GPS data
  to the emergency coordination center \cite{mobileApp}. The drone is not equipped with parachute or rope systems to supply the AED, thus landing at the patient's site is mandatory.

  \subsection{Analysis}
  We consider a test instance that consists of $104$ base stations and $1500$ patients' sites
  that have been randomly sampled from points on the hiking trail network.
  The label $O_{\overline{t}}$ denotes instances for which inequality \eqref{eq:minfeasibletime} does not hold, i.e. no feasible solution can be found.
  When we choose to minimize the number of drones, i.e. the travel time is not part of the objective function,
  we calculate the travel time solely based on the selected base stations. Hence, we ignore the
  assignment variables $x_{ij}$ and choose the assignment of patients to
  base stations  $\mathcal{B}^* \subseteq \mathcal{B}$ having a drone
  assigned to
  based on the travel time, i.e. for each patient $p \in\mathcal{P}$ the corresponding
  base station is determined as $\argmin_{b \in \mathcal{B}^*} t(b,p)$.

  \subsubsection{Optimal Allocation of Defibrillator Drones - Minimizing Average Travel Times and Number of Drones.}
  \paragraph{Setup.}
  In this first scenario we choose the \textit{minimal average travel time} as
  objective function, i.e., $\alpha=0$.
  We vary the maximal number of drones in steps of $1$ and
  report the results in Table \ref{tab:minTime}. From interval $40$ to $100$ we abstract the results in steps of $10$ as no remarkable changes in values can be observed within this range.

\begin{figure}[!ht]
    \begin{center}
      \includegraphics[width=\textwidth]{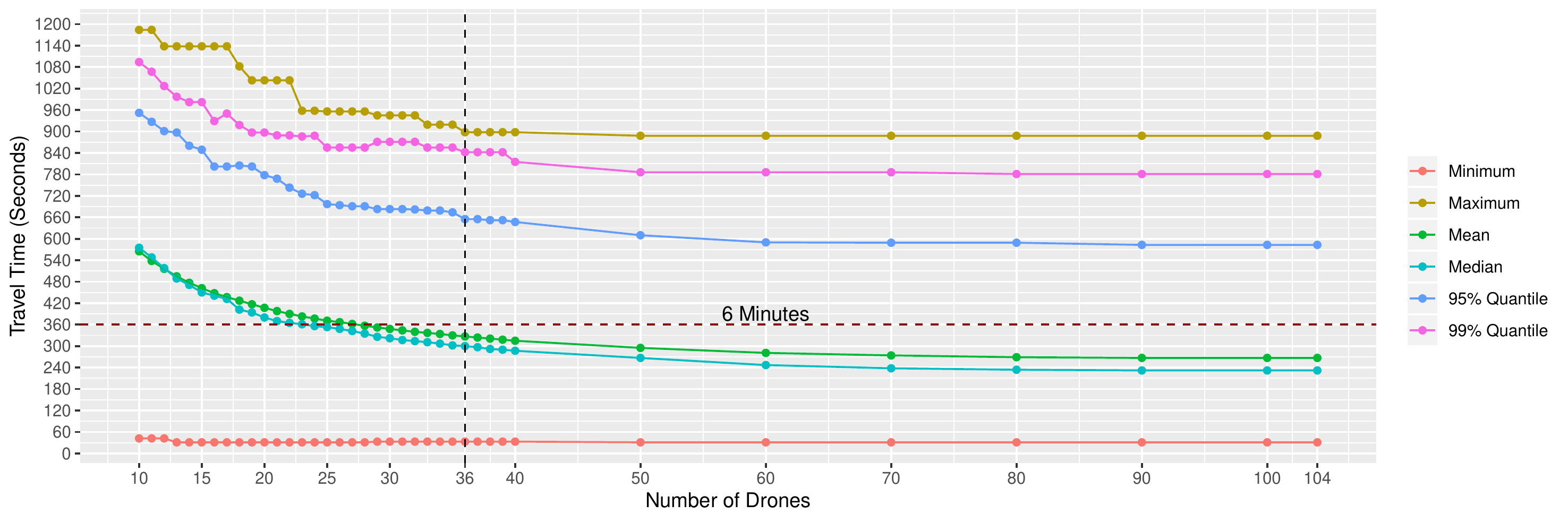}
    \end{center}
    \caption{Plot of Table    \ref{tab:minTime}.
    The x-axis indicates the number of available drones $s$. For each value
    the minimal and maximal travel time between a selected base station and an assigned patient is reported.
    Moreover, the mean and median such as the  $95 \, \%$ and the $99 \, \%$ quantiles are reported.}
    \label{fig:tableMinTime}
  \end{figure}

\begin{figure}[!ht]
    \includegraphics[width=0.45 \textwidth]{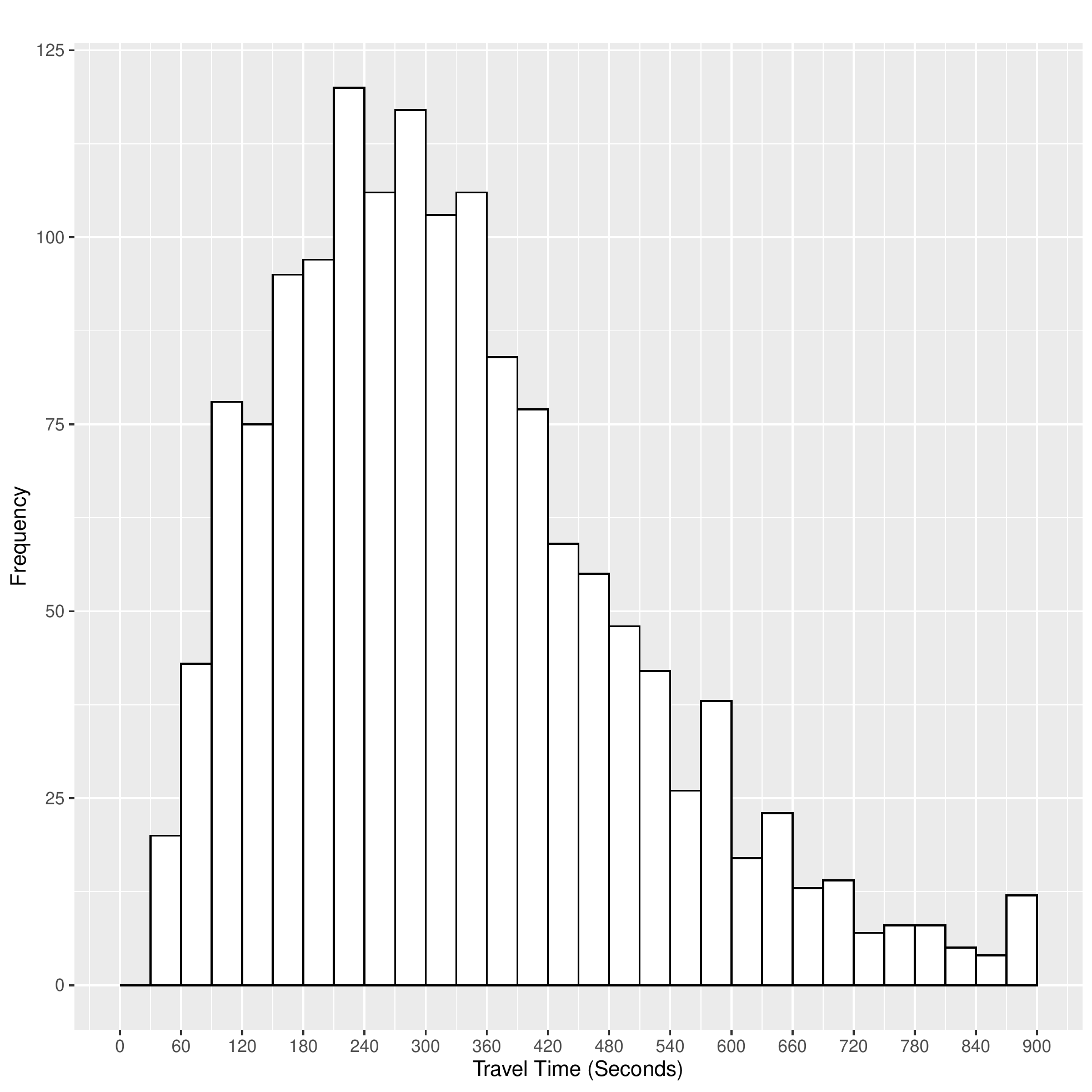}
    \hfill
    \includegraphics[width=0.45 \textwidth]{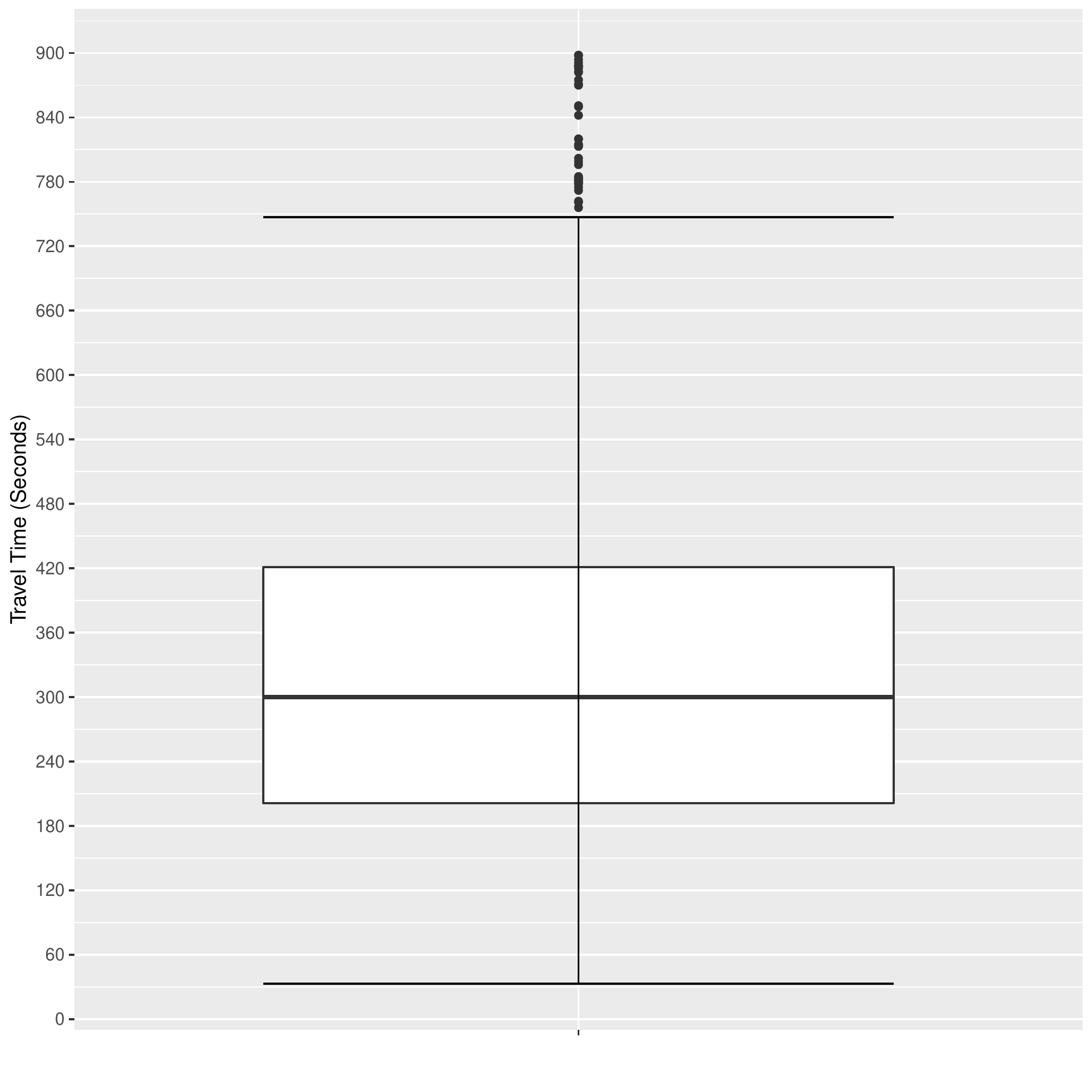}
    \caption{Histogram and Boxplot of the travel times in
    the solution of configuration \texttt{ATTs36}.
    }
    \label{fig:stats}
  \end{figure}

  \paragraph{Results.}
  We plot the results from Table \ref{tab:minTime} in  Figure \ref{fig:tableMinTime}.
  Hence, we choose  $s=36$ as the preferable network configuration of drone base
  stations, as the mean travel time of drones ($05 \colon 27$) is well below the $6$-minute threshold in which early defibrillation should be provided \cite{nichol1999cumulative}. Further, there are no remarkable changes of the $95 \, \%$ quantile values
  for $s>36$. We refer to this configuration as \texttt{ATTs36} (average travel time with $s=36$) and plot the corresponding data in Figure \ref{fig:stats}.
  This allocation of drone base stations in the region allows travel times
  of drones to the patient's sites in an average time of
  $05\colon 27$. Moreover, $50 \,\%$ of all patients can be supplied with an AED within $05\colon00$, which
  translates to a rate of survival between $50 \, \% $ and $ 70 \, \%$ if CPR is immediately provided.
  Further, $95 \,\%$ of the patients can be reached by an AED drone within $10\colon55$.
  However, such long travel times are no more beneficial to the patients, keeping in mind the decrease of survival rate of 8-16\% per minute without CPR and AED.
  The $99 \, \%$ quantile of $14\colon02$ further underlines the insufficient
  performance of the
  subsequent drone type in \texttt{ATTs36} and puts into question the benefit of this response system to the patient's survival.
  We illustrate the configuration in Figure \ref{fig:mapd36slow} (included in the Appendix).
  Patients that can be reached in $06\colon 00$ or less are indicated in blue,
  patients that can be reached between $06\colon 00$ and $11\colon 00$ are
  indicated in green, and patients that can not be reached in less than $11\colon 00$ are
  indicated in red. Potential base stations are represented by using brown circles and
  red cross symbols denote selected base stations according to the optimal
  solution of the ILP. We also illustrate the hiking trails in the region by brown lines.
  Analysing the configuration in Figure \ref{fig:mapd36slow} reveals an accumulation of
  patients with travel times higher than $11\colon 00$ (colored in red) located
  in the most eastern part of Val Venosta. The reason for this is the low availability
  of potential base stations (i.e. shelter huts) in this region. Hence, it becomes
  obvious that travel times of drones in the analyzed setting can only be
  accelerated by using faster drones due to the unavailability of additional base stations in the region.

  \begin{table}[!ht]
    \footnotesize
    \centering
    \begin{tabular}{ r | r | r | r | r | r | r }
      \multicolumn{7}{l}{\textbf{Minimizing Average Travel Time:}  \qquad  $t_{max}= 20 \colon 00$ }\\
      \hline
      $s$ & min($t$) & max($t$) & mean($t$) & median($t$) & 95 \% & 99 \% \\
      \hline \hline
      $\leq 9$ & $O_{\overline{t}}$ & $O_{\overline{t}}$  & $O_{\overline{t}}$  &  $O_{\overline{t}} $ & $O_{\overline{t}}$  & $O_{\overline{t}}$ \\
      $10$ & $00\colon 42 $ & $19\colon44 $ & $09\colon 25 $ & $ 09\colon35 $ & $15\colon52$ & $18\colon14$ \\
      $11$ & $00\colon 42 $ & $19\colon44 $ & $08\colon 58 $ & $ 09\colon08 $ & $15\colon27$ & $17\colon47$ \\
      $12$ & $00\colon 42 $ & $18\colon58 $ & $08\colon 36 $ & $ 08\colon38 $ & $15\colon01$ & $17\colon07$ \\
      $13$ & $00\colon 31 $ & $18\colon58 $ & $08\colon 15 $ & $ 08\colon09 $ & $14\colon57$ & $16\colon37$ \\
      $14$ & $00\colon 31 $ & $18\colon58 $ & $07\colon 57 $ & $ 07\colon51 $ & $14\colon20$ & $16\colon22$ \\
      $15$ & $00\colon 31 $ & $18\colon58 $ & $07\colon 42 $ & $ 07\colon30 $ & $14\colon09$ & $16\colon22$ \\
      $16$ & $00\colon 31 $ & $18\colon58 $ & $07\colon 28 $ & $ 07\colon21 $ & $13\colon22$ & $15\colon29$ \\
      $17$ & $00\colon 31 $ & $18\colon58 $ & $07\colon 17 $ & $ 07\colon12 $ & $13\colon22$ & $15\colon50$ \\
      $18$ & $00\colon 31 $ & $18\colon02 $ & $07\colon 07 $ & $ 06\colon42 $ & $13\colon25$ & $15\colon18$ \\
      $19$ & $00\colon 31 $ & $17\colon23 $ & $06\colon 57 $ & $ 06\colon34 $ & $13\colon22$ & $14\colon57$ \\
      $20$ & $00\colon 31 $ & $17\colon23 $ & $06\colon 47 $ & $ 06\colon20 $ & $12\colon58$ & $14\colon57$ \\
      $21$ & $00\colon 31 $ & $17\colon23 $ & $06\colon 38 $ & $ 06\colon10 $ & $12\colon48$ & $14\colon49$ \\
      $22$ & $00\colon 31 $ & $17\colon23 $ & $06\colon 30 $ & $ 06\colon05 $ & $12\colon23$ & $14\colon49$ \\
      $23$ & $00\colon 31 $ & $15\colon58 $ & $06\colon 23 $ & $ 06\colon01 $ & $12\colon06$ & $14\colon46$ \\
      $24$ & $00\colon 31 $ & $15\colon58 $ & $06\colon 17 $ & $ 05\colon56 $ & $12\colon02$ & $14\colon48$ \\
      $25$ & $00\colon 31 $ & $15\colon56 $ & $06\colon 11 $ & $ 05\colon53 $ & $11\colon37$ & $14\colon15$ \\
      $26$ & $00\colon 31 $ & $15\colon56 $ & $06\colon 07 $ & $ 05\colon48 $ & $11\colon34$ & $14\colon15$ \\
      $27$ & $00\colon 31 $ & $15\colon56 $ & $06\colon 02 $ & $ 05\colon42 $ & $11\colon31$ & $14\colon15$ \\
      $28$ & $00\colon 31 $ & $15\colon56 $ & $05\colon 57 $ & $ 05\colon35 $ & $11\colon31$ & $14\colon15$ \\
      $29$ & $00\colon 33 $ & $15\colon45 $ & $05\colon 52 $ & $ 05\colon26 $ & $11\colon23$ & $14\colon31$ \\
      $30$ & $00\colon 33 $ & $15\colon45 $ & $05\colon 48 $ & $ 05\colon22 $ & $11\colon23$ & $14\colon31$ \\
      $31$ & $00\colon 33 $ & $15\colon45 $ & $05\colon 44 $ & $ 05\colon17 $ & $11\colon23$ & $14\colon31$ \\
      $32$ & $00\colon 33 $ & $15\colon45 $ & $05\colon 40 $ & $ 05\colon14 $ & $11\colon22$ & $14\colon31$ \\
      $33$ & $00\colon 33 $ & $15\colon19 $ & $05\colon 37 $ & $ 05\colon11 $ & $11\colon19$ & $14\colon15$ \\
      $34$ & $00\colon 33 $ & $15\colon19 $ & $05\colon 34 $ & $ 05\colon07 $ & $11\colon19$ & $14\colon15$ \\
      $35$ & $00\colon 33 $ & $15\colon19 $ & $05\colon 30 $ & $ 05\colon02 $ & $11\colon14$ & $14\colon15$ \\
      $36$ & $00\colon 33 $ & $14\colon58 $ & $05\colon 27 $ & $ 05\colon00 $ & $10\colon55$ & $14\colon02$ \\
      $37$ & $00\colon 33 $ & $14\colon58 $ & $05\colon 24 $ & $ 04\colon57 $ & $10\colon55$ & $14\colon02$ \\
      $38$ & $00\colon 33 $ & $14\colon58 $ & $05\colon 21 $ & $ 04\colon52 $ & $10\colon52$ & $14\colon02$ \\
      $39$ & $00\colon 33 $ & $14\colon58 $ & $05\colon 18 $ & $ 04\colon50 $ & $10\colon52$ & $14\colon02$ \\
      $40$ & $00\colon 33 $ & $14\colon58 $ & $05\colon 15 $ & $ 04\colon47 $ & $10\colon47$ & $13\colon35$ \\
      $50$ & $00\colon 31 $ & $14\colon48 $ & $04\colon 55 $ & $ 04\colon27 $ & $10\colon10$ & $13\colon06$ \\
      $60$ & $00\colon 31 $ & $14\colon48 $ & $04\colon 41 $ & $ 04\colon07 $ & $09\colon 50$ & $13\colon06$ \\
      $70$ & $00\colon 31 $ & $14\colon48 $ & $04\colon 34 $ & $ 03\colon58 $ & $09\colon 49$ & $13\colon06$ \\
      $80$ & $00\colon 31 $ & $14\colon48 $ & $04\colon 29 $ & $ 03\colon54 $ & $09\colon 49$ & $13\colon01$ \\
      $90$ & $00\colon 31 $ & $14\colon48 $ & $04\colon 27 $ & $ 03\colon52 $ & $09\colon 43$ & $13\colon01$ \\
      $100$ & $00\colon 31 $ & $14\colon48 $ & $04\colon 27 $ & $ 03\colon52 $ & $09\colon 43$ & $13\colon01$ \\
      $104$ & $00\colon 31 $ & $14\colon48 $ & $04\colon 27 $ & $ 03\colon52 $ & $09\colon 43$ & $13\colon01$\\
      \hline
    \end{tabular}
    \caption{Table summarizing the generated results when minimizing the average travel time using the \textit{LifeDrone AED} system (mm:ss).}
    \label{tab:minTime}
  \end{table}

  \clearpage

  \paragraph{Setup - Faster Drone.}
  As the considered drone model \textit{LifeDrone AED} has rather low vertical and horizontal speed, which is a major drawback when large obstacles like mountains must be
  overcome, we compare it against a faster drone model.
  Therefore we consider the \textit{Wingcopter 178} drone, which is
  commercially available and capable of carrying an AED of
  type \textit{Philips Heartstart}. This drone model has parameters of $v_{vert}^{+}=6 \, m/s$, $v_{vert}^{-}=6 \, m/s$, and
  $v_{hor}=36.1 \,m/s$ and $c_{start}=20 \,s$ \cite{Wingcopter}. In Table \ref{tab:droneParameters} we summarize the parameters of both drone models considered in this work.

  \begin{table}[!ht]
    \begin{center}
      \begin{tabular}{ l| l |l |l }
        Parameter  &  Description & \textit{LifeDrone AED}  & \textit{Wingcopter 178} \\
        \hline \hline
        $v_{vert}^{+}$ & vertical ascending  speed   & $2.5 \, m/s$  &  $6 \, m/s$ \\
        $v_{vert}^{-}$ &    vertical descending  speed & $2.5 \, m/s$  & $6 \, m/s$ \\
        $v_{hor}$ &   horizontal travel speed &  $17.9 \,m/s$  & $36.1 \, m/s$ \\
        $c_{start}$  & start-up time & $30 \,s$ & $20 \,s$  \\
        & maximal wind speed to operate & $12 \, m/s$ & $15 \, m/s$  \\
        \hline
      \end{tabular}
      \caption{Summary of the parameters of the considered drone models.}
      \label{tab:droneParameters}
    \end{center}
  \end{table}

  \paragraph{Results - Faster Drone.}We compare the results of the slower drone against the faster drone in \texttt{ATTs36}. Using the faster drone in this setting shows remarkable improvements in terms of travel times.
  Firstly, $\overline{t}$ reduces to $06\colon 41$.
  Moreover, the average travel time with the faster drone type amounts to $02\colon 36$, which is an improvement of $02\colon 51$ compared to the slower one. Moreover, $50 \, \%$ of all patients (i.e. median value) can be reached within $02\colon 24$, which is a reduction in time of $02\colon 36$. Hence, the faster drone reaches $95 \,\%$ of the patients within $05\colon 00$ (compared to $10\colon55$ using the slower drone). Improvements can also be observed with the $99 \, \%$ quantile that amounts to $06\colon 23$ , which indicates a time reduction of $07\colon 39$ compared to the slower drone.
  In Figure \ref{fig:mapd36fast} we again illustrate the selected base stations. When comparing Figure \ref{fig:mapd36slow} and Figure \ref{fig:mapd36fast}, i.e. demand covered by \textit{LifeDrone AED} and \textit{Wingcopter 178}, a drastic reduction of patients with travel times higher than $11\colon 00$ (patients colored in red in Figure \ref{fig:mapd36slow} turned blue in Figure \ref{fig:mapd36fast}) can be observed. In conclusion, using the same number of drones but considering a faster drone type can be beneficial for patients even in remotely located areas where no potential base station is available.\\
  \begin{table}[!ht]
    \footnotesize
    \centering
    \begin{tabular}{ l | r | r | r | r | r | r }
      \multicolumn{7}{l}{\textbf{Minimizing Average Travel Time:}  \qquad  $t_{max}= 20 \colon 00$ } \\
      \hline
      $s$ & min($t$) & max($t$) & mean($t$) & median($t$) & 95 \% & 99 \% \\
      \hline \hline
      $01$ & $00\colon46 $ & $19\colon48 $ & $11\colon32 $ & $12\colon04 $ & $17\colon37$ & $19\colon12$ \\
      $02$ & $00\colon56 $ & $17\colon00 $ & $08\colon40 $ & $08\colon21 $ & $14\colon24$ & $15\colon30$ \\
      $03$ & $00\colon24 $ & $16\colon00 $ & $07\colon22 $ & $07\colon27 $ & $12\colon07$ & $13\colon53$ \\
      $04$ & $00\colon29 $ & $13\colon14 $ & $06\colon28 $ & $06\colon23 $ & $10\colon30$ & $12\colon00$ \\
      $05$ & $00\colon26 $ & $12\colon07 $ & $05\colon44 $ & $05\colon40 $ & $09\colon28$ & $10\colon31$ \\
      $06$ & $00\colon24 $ & $12\colon12 $ & $05\colon19 $ & $05\colon15 $ & $09\colon15$ & $10\colon42$ \\
      $07$ & $00\colon24 $ & $12\colon12 $ & $04\colon60 $ & $04\colon56 $ & $08\colon46$ & $10\colon11$ \\
      $08$ & $00\colon24 $ & $10\colon38 $ & $04\colon43 $ & $04\colon39 $ & $08\colon22$ & $09\colon52$ \\
      $09$ & $00\colon20 $ & $10\colon36 $ & $04\colon29 $ & $04\colon23 $ & $07\colon57$ & $09\colon52$ \\
      $10$ & $00\colon20 $ & $10\colon36 $ & $04\colon17 $ & $04\colon20 $ & $07\colon16$ & $08\colon35$ \\
      $11$ & $00\colon20 $ & $10\colon36 $ & $04\colon17 $ & $04\colon20 $ & $07\colon16$ & $08\colon35$ \\
      $12$ & $00\colon20 $ & $10\colon36 $ & $03\colon56 $ & $03\colon52 $ & $07\colon00$ & $07\colon48$ \\
      $13$ & $00\colon20 $ & $10\colon36 $ & $03\colon48 $ & $03\colon36 $ & $06\colon54$ & $07\colon45$ \\
      $14$ & $00\colon20 $ & $10\colon36 $ & $03\colon40 $ & $03\colon29 $ & $06\colon34$ & $07\colon36$\\
      $15$ & $00\colon20 $ & $10\colon36 $ & $03\colon34 $ & $03\colon29 $ & $06\colon15$ & $07\colon36$\\
      $16$ & $00\colon20 $ & $10\colon36 $ & $03\colon29 $ & $03\colon24 $ & $06\colon09$ & $07\colon36$ \\
      $17$ & $00\colon20 $ & $10\colon36 $ & $03\colon24 $ & $03\colon12 $ & $06\colon15$ & $07\colon34$ \\
      $18$ & $00\colon20 $ & $10\colon36 $ & $03\colon24 $ & $03\colon12 $ & $06\colon15$ & $07\colon34$ \\
      $19$ & $00\colon20 $ & $10\colon36 $ & $03\colon15 $ & $03\colon02 $ & $06\colon07$ & $07\colon14$ \\
      $20$ & $00\colon20 $ & $10\colon36 $ & $03\colon11 $ & $02\colon60 $ & $05\colon59$ & $07\colon04$ \\
      $21$ & $00\colon20 $ & $10\colon36 $ & $03\colon07 $ & $02\colon56 $ & $05\colon51$ & $07\colon04$ \\
      $22$ & $00\colon20 $ & $10\colon36 $ & $03\colon04 $ & $02\colon54 $ & $05\colon40$ & $07\colon00$ \\
      $23$ & $00\colon20 $ & $10\colon36 $ & $03\colon01 $ & $02\colon52 $ & $05\colon39$ & $06\colon57$ \\
      $24$ & $00\colon20 $ & $10\colon36 $ & $02\colon59 $ & $02\colon48 $ & $05\colon34$ & $06\colon54$ \\
      $25$ & $00\colon20 $ & $10\colon36 $ & $02\colon56 $ & $02\colon47 $ & $05\colon23$ & $06\colon41$ \\
      $26$ & $00\colon20 $ & $10\colon36 $ & $02\colon54 $ & $02\colon45 $ & $05\colon23$ & $06\colon41$ \\
      $27$ & $00\colon20 $ & $10\colon36 $ & $02\colon52 $ & $02\colon43 $ & $05\colon24$ & $06\colon42$ \\
      $28$ & $00\colon20 $ & $07\colon14 $ & $02\colon50 $ & $02\colon43 $ & $05\colon18$ & $06\colon23$ \\
      $29$ & $00\colon22 $ & $06\colon56 $ & $02\colon48 $ & $02\colon39 $ & $05\colon13$ & $06\colon34$ \\
      $30$ & $00\colon22 $ & $06\colon56 $ & $02\colon46 $ & $02\colon35 $ & $05\colon12$ & $06\colon34$ \\
      $31$ & $00\colon22 $ & $06\colon56 $ & $02\colon44 $ & $02\colon32 $ & $05\colon12$ & $06\colon34$ \\
      $32$ & $00\colon22 $ & $06\colon56 $ & $02\colon42 $ & $02\colon31 $ & $05\colon12$ & $06\colon34$ \\
      $33$ & $00\colon22 $ & $06\colon56 $ & $02\colon41 $ & $02\colon28 $ & $05\colon12$ & $06\colon34$ \\
      $34$ & $00\colon22 $ & $06\colon56 $ & $02\colon39 $ & $02\colon26 $ & $05\colon12$ & $06\colon34$ \\
      $35$ & $00\colon22 $ & $06\colon56 $ & $02\colon37 $ & $02\colon25 $ & $05\colon07$ & $06\colon34$ \\
      $36$ & $00\colon22 $ & $06\colon56 $ & $02\colon36 $ & $02\colon24 $ & $05\colon00$ & $06\colon23$ \\
      $37$ & $00\colon21 $ & $06\colon56 $ & $02\colon34 $ & $02\colon22 $ & $05\colon00$ & $06\colon23$ \\
      $38$ & $00\colon21 $ & $06\colon51 $ & $02\colon33 $ & $02\colon20 $ & $04\colon58$ & $06\colon17$ \\
      $39$ & $00\colon21 $ & $06\colon51 $ & $02\colon32 $ & $02\colon20 $ & $04\colon58$ & $06\colon17$ \\
      $40$ & $00\colon21 $ & $06\colon51 $ & $02\colon30 $ & $02\colon19 $ & $04\colon54$ & $06\colon17$ \\
      $50$ & $00\colon20 $ & $06\colon41 $ & $02\colon21 $ & $02\colon08 $ & $04\colon42$ & $06\colon07$ \\
      $60$ & $00\colon20 $ & $06\colon41 $ & $02\colon15 $ & $01\colon59 $ & $04\colon33$ & $06\colon07$ \\
      $70$ & $00\colon20 $ & $06\colon41 $ & $02\colon11 $ & $01\colon56 $ & $04\colon29$ & $06\colon07$ \\
      $80$ & $00\colon20 $ & $06\colon41 $ & $02\colon09 $ & $01\colon54 $ & $04\colon28$ & $06\colon07$ \\
      $90$ & $00\colon20 $ & $06\colon41 $ & $02\colon08 $ & $01\colon52 $ & $04\colon28$ & $06\colon07$ \\
      $100$ & $00\colon20 $ & $06\colon41 $ & $02\colon08 $ & $01\colon52 $ & $04\colon28$ & $06\colon07$ \\
      $104$ & $00\colon20 $ & $06\colon41 $ & $02\colon08 $ & $01\colon52 $ & $04\colon28$ & $06\colon07$ \\
      \hline
    \end{tabular}

    \caption{Table summarizing the generated results when minimizing the average travel time using the 1 using the \textit{Wingcopter 178} system  ($\text{mm}\colon \text{ss}$).}
    \label{tab:minTimefast}
  \end{table}
  \par
  Alternatively we considered to minimize the number of used drones
  while $t_{max}$ is fixed, i.e. $\alpha=1$.
  Considering the \textit{LifeDrone AED} and \textit{Wingcopter 178}, we vary the maximal allowed time $t_{max}$ in steps of 15 seconds, starting at $\overline{t}=14\colon 48$ and $\overline{t}=06\colon 41$, respectively.
  The large value of $\overline{t}$ implies a rather large $t_{max}$. Together with the fact that
  the travel time is not part of the objective, this scenario yields unacceptable results.
  However, the authors decided not to consider objective functions where
  the number of drones is weighted against the travel times due to ethical reasons.
  The results are reported in Table \ref{tab:minDrones} and Table \ref{tab:minDronesfast},
  respectively, which can be found in the Appendix  of the paper.

  \paragraph{Summary of the Process (Pseudo code).}
  We summarize the process (for the \textit{LifeDrone AED}) of generating our results as following:
  \begin{enumerate}
    \item Collect and prepare the data as described in Subsection \ref{subsec:dataprep}.
    \item Repeatedly solve the problem instance with changing parameters. \\
    - For $t_{\max}=20\colon 00$ $\alpha=0$, solve the instance
    for $s=1,\ldots, 104$.
    \item Analyze the resulting drone travel times.
  \end{enumerate}

  \subsubsection{Optimal Allocation of Defibrillator Drones with Backup}

  \paragraph{Setup - Faster Drone with Backup.}
  In this subsection, we discuss the unlikely event of having two SCA patients located
  close to each other. In this case, both SCA patients would be served from the same base station, which is clearly not possible, as the responding drone cannot visit both patients' sites.
  Consequently, a defibrillator drone from another base station must be sent to one of the patients' sites. The use of a backup drone is a reasonable choice to overcome this kind of situation.
  Hence, for each patient there are two drones assigned.
  By default,
  the closest drone, denoted as drone 1, responds to the patient.
  In case that drone 1 is not available, the second closest drone,
  denoted as drone 2, is sent to the patient's site.
  In this work, we consider the following two approaches of assigning drone 2 to the patients:
  \begin{itemize}
    \item
    \texttt{B1}:
    In this case, we stick to the original assumption that at most one drone can be assigned to each base station. Hence, drone 2 is located at a different base station than drone 1.
    Consequently, drone 2 is sent from the second nearest
    base station,
    i.e. we  change Constraints \eqref{ineq:eachPatient}  to
    $$\sum_{i\in [m]} x_{ij}  \geq 2,   \quad   j \in [q].$$

    \item
    \texttt{B2}:
    Here, we consider the case that a base station can be equipped
    with up to two drones.
    Now, for a given patient, there are two options. Either drone 2 is located at the same base station as drone 1, or
    drone 2 is located at a different base station.
      In order compute this scenario using   Model  \eqref{form:fone},
    we extend the problem instance such that each potential base station is duplicated.
  \end{itemize}

  \paragraph{Results - Faster Drone with Backup.}

  Again, we consider the \textit{Wingcopter 178} using $s=36$ drones and set $t_{max}= 20 \colon 00$.
  We compare the results against the setup without backup (\texttt{B0}) and summarize
  the results in Table \ref{tab:backup}. In doing so, we report the values for drone 1 and drone 2.
  For \texttt{B1} and \texttt{B2}, we calculated the differences between the response times of drone 1 and drone 2 for each patient and
  we report the statistics in the table.

  \begin{table}[!ht]
    \footnotesize
    \centering
    \begin{tabular}{l|  r | r | r | r | r | r }
        & min($t$) & max($t$) & mean($t$) & median($t$) & 95 \% & 99 \% \\
      \hline\hline
      \texttt{B0}  & $0:22 $ & $6:56 $ & $2:36 $ & $ 2:24 $ & $5:00$ & $6:23$ \\

      \hline
      \texttt{B1} &&&&&& \\
      \quad Drone 1
      & $0:20 $ & $10:36 $ & $2:43 $ & $ 2:27 $ & $5:33$ & $6:53$ \\
      \quad Drone 2
      & $1:18 $ & $10:53 $ & $4:02 $ & $ 3:53 $ & $6:23$ & $7:32$ \\
      \quad Difference &
      $0:01 $ & $5:37 $ & $1:19 $ & $ 1:04 $ & $3:17$ & $3:56$\\
      \hline
      \texttt{B2} &&&&&& \\
      \quad Drone 1
      & $0:20 $ & $10:36 $ & $3:13 $ & $ 3:01 $ & $6:07$ & $7:14$\\
      \quad Drone 2
      & $0:20 $ & $10:36 $ & $3:26 $ & $ 3:15 $ & $6:15$ & $7:34$ \\
      \quad Difference
      &  $0:00 $ & $5:52 $ & $0:12 $ & $ 0:00 $ & $1:56$ & $3:53$
      \\

      \multicolumn{7}{c}{16 base stations have two drones allocated} \\
      \hline
    \end{tabular}
    \caption{Table summarizing the comparison of the \textit{Wingcopter 178} system.
    }
    \label{tab:backup}
  \end{table}
  Analyzing the results shows that there is virtually no increase in the response time for
  drone 1 in \texttt{B1} (except for outliers). However, drone 2 arrives notably later, but for
  $95 \, \%$ of the patients the use of drone 2 is still beneficial.
  Moreover, we notice that the response times of drone 1 in \texttt{B2} is longer than for  drone 1 in \texttt{B1}.
  In \texttt{B1} drone 2 arrives on average $01 \colon 19$ later at the same patient's site than drone 1. In \texttt{B2} this difference amounts to only $00 \colon 12$. Also the $95 \, \%$ quantile in \texttt{B2} indicates that the patients still benefit from the AED delivery.
  Note that in \texttt{B2} we use $16$ base stations equipped with $2$ drones and $4$ base stations equipped with $1$ drone.
  Further, the slight decrease in the minimal response time of drone 1 in
  \texttt{B1} and \texttt{B2} (compared to \texttt{B0})
  can be explained by the model choosing different assignments of drones to base stations.
  This is a valid behavior as the objective function is to minimize the average travel time.

  To sum up, this analysis shows that
  considering backups in the assignment of drones to
  potential base stations enables a timely AED delivery even in the unlikely situation of simultaneously
  having two SCA patients closely located to each other.

  \subsubsection{Travel Time Comparison of Defibrillator Drones and Air Ambulance}
  \paragraph{Setup.}

  In order to compare the performance of the defibrillator drone system to the conventional air ambulance, we first give a short overview of the air ambulance system in South Tyrol. In the region there are three ambulance helicopters available that are located in Bozen, Brixen and Gröden (highlighted in Figure \ref{fig:helis}). The latter is only available in the summer season while the others offer all-season response \cite{heli3}. As previously stated, the base stations of the helicopters are geographically decoupled from Val Venosta leading to relatively long flight times to locations in this region. Interviews with experts from the mountain rescue service of South Tyrol underline this by pointing to flight times varying between $14\colon 00$ and $25\colon 00$ to locations in the most western parts of Val Venosta.
  Consequently, we argue that an optimally allocated fleet of drones could assist in reducing the time span between early defibrillation and helicopter arrival in case of SCA in the region.
  In order to learn more about helicopter flight times and to validate the expert statements we set up a meeting with representatives of the local air ambulance provider in January 2019. The mountain rescuers' statements were confirmed and further insights into air ambulance service in South Tyrol were gathered. In addition, a data set including 100 flight times of the three ambulance helicopters to historical emergencies (and corresponding latitude and longitude data of the emergency locations) in Val Venosta in 2018 was provided. The data set comprises information on the responding helicopter, its departure time and arrival time on the scene. It is to be noted that the data set does not include patient related information, i.e. it is not reported what kind of medical emergency occurred. The data set validated the assumption of having most patients located on official hiking trails. Therefore, the data can be used to compare the travel times of the helicopters against the drone network. Descriptive analysis of the data set reveals flight times with a minimum of $17\colon 00$, maximum of $48\colon 00$, and mean of $26\colon35$, which further supports the expert statements.
  For comparison we use the given defibrillator drone network from \texttt{ATTs36} using \textit{Wingcopter 178} without backup supply. Consequently, we determine the shortest travel time for each patient based on the selected base stations and compare the generated results with the historical flight times of the data set.

  \paragraph{Results.}
  We report on the results of travel times of defibrillator drones against helicopters responding to historical incidents from the data set discussed before. Analyzing the generated travel times of defibrillator drones shows that flight times to all emergency locations are well below 6 minutes
  except for one patient. The helicopter cannot undercut this threshold in any case. The minimal travel time for defibrillator drones amounts to $00\colon 22$ and for the helicopter it is $17\colon 00$. The extremely short flight time for the defibrillator drone with only $00\colon 22$ is achieved by a drone departing from a fire rescue base station located really close to a patient's location. The maximum travel time for the drone system amounts to $06\colon 09$, while it is $48\colon 00$ with the helicopter. The average flight time of defibrillator drones to patients' sites is  $02\colon 05$ and $26\colon 35$ for helicopters. The defibrillator drones reach $95 \,\%$ of the patients within $03\colon39$ (compared to $36\colon56$ with the helicopters). The $99 \, \%$ quantile for the defibrillator drones amounts to $05\colon01$ and $47\colon00$
  for the helicopters.
  Interesting to observe is that in the worst case, the defibrillator drone requires $15.56 \, \%$ of the flight time of the helicopter (see Table \ref{tab:helicoptercomparison}). We consider patients' locations in Table \ref{tab:helicoptercomparison}, where the flight time of the drone is greater than $10\, \%$ of the corresponding helicopter flight time. Additionally, we plot these locations in Figure \ref{fig:maphelidrone} (included in the Appendix). All in all, we can observe remarkable reductions of the response times using the defibrillator drones compared against the existing helicopter fleet in the region.

  \begin{table}[!ht]
    \footnotesize
    \centering
    \begin{tabular}{  r | r | r  }
      $t_{D}$ & $t_{H}$ & $t_{D} / t_{H}  (\%) $ \\
      \hline\hline
      $01\colon54$ & $19\colon00$ & $10.00\, \% $ \\
      $02\colon12$ & $22\colon00$ & $10.00\, \% $ \\
      $02\colon25$ & $24\colon00$ & $10.07\, \% $ \\
      $02\colon19$ & $23\colon00$ & $10.07\, \% $ \\
      $01\colon56$ & $19\colon00$ & $10.18\, \% $ \\
      $02\colon52$ & $28\colon00$ & $10.24\, \% $ \\
      $02\colon22$ & $23\colon00$ & $10.29\, \% $ \\
      $02\colon55$ & $28\colon00$ & $10.42\, \% $ \\
      $02\colon37$ & $25\colon00$ & $10.47\, \% $ \\
      $02\colon22$ & $22\colon00$ & $10.76\, \% $ \\
      $02\colon56$ & $27\colon00$ & $10.86\, \% $ \\
      $02\colon30$ & $23\colon00$ & $10.87\, \% $ \\
      $02\colon30$ & $23\colon00$ & $10.87\, \% $ \\
      $02\colon57$ & $27\colon00$ & $10.93\, \% $ \\
      $02\colon34$ & $23\colon00$ & $11.16\, \% $ \\
      $03\colon02$ & $27\colon00$ & $11.23\, \% $ \\
      $02\colon16$ & $20\colon00$ & $11.33\, \% $ \\
      $02\colon47$ & $24\colon00$ & $11.60\, \% $ \\
      $03\colon39$ & $31\colon00$ & $11.77\, \% $ \\
      $02\colon50$ & $24\colon00$ & $11.81\, \% $ \\
      $02\colon22$ & $20\colon00$ & $11.83\, \% $ \\
      $03\colon22$ & $28\colon00$ & $12.02\, \% $ \\
      $03\colon12$ & $25\colon00$ & $12.80\, \% $ \\
      $04\colon14$ & $33\colon00$ & $12.83\, \% $ \\
      $03\colon14$ & $25\colon00$ & $12.93\, \% $ \\
      $02\colon22$ & $18\colon00$ & $13.15\, \% $ \\
      $02\colon30$ & $19\colon00$ & $13.16\, \% $ \\
      $02\colon25$ & $18\colon00$ & $13.43\, \% $ \\
      $02\colon36$ & $19\colon00$ & $13.68\, \% $ \\
      $03\colon27$ & $25\colon00$ & $13.80\, \% $ \\
      $02\colon25$ & $17\colon00$ & $14.22\, \% $ \\
      $05\colon00$ & $35\colon00$ & $14.29\, \% $ \\
      $02\colon44$ & $19\colon00$ & $14.39\, \% $ \\
      $03\colon55$ & $27\colon00$ & $14.51\, \% $ \\
      $03\colon39$ & $25\colon00$ & $14.60\, \% $ \\
      $06\colon09$ & $42\colon00$ & $14.64\, \% $ \\
      $03\colon34$ & $24\colon00$ & $14.86\, \% $ \\
      $03\colon34$ & $23\colon00$ & $15.51\, \% $ \\
      $02\colon49$ & $18\colon00$ & $15.65\, \% $ \\
      \hline
    \end{tabular}

    \caption{Table summarizing the comparison of the \textit{Wingcopter 178} system in the \texttt{ATTs36} configuration against conventional helicopter rescue.
    We report the travel time  ($\text{mm}\colon \text{ss}$) of the defibrillator drone network $t_{D}$,
    the travel time of the helicopter $t_{H}$,
    and the relative travel time of the defibrillator drones $t_{D} / t_{H} $ compared to the helicopter rescue.
    }
    \label{tab:helicoptercomparison}
  \end{table}

  \section{Discussion \& Conclusion} \label{sec:concl}
  Using our optimization model, we showed that (1) a dense network of drones allocated
  in a remote region like the Alps in South Tyrol could deliver AEDs to emergency
  patients on hiking paths, (2) the median travel time interval of the drones
  is  $02\colon 24$ (using the \textit{Wingcopter 178}), (3) that an optimally allocated
  network of defibrillator drones could reduce the time span between AED provision
  and helicopter arrival, and (4) this substantially reduced time interval
  may be associated with a beneficial outcome of SCA, namely a vastly improved chance of survival. Besides, the results
  show that considering backups in the assignment of defibrillator drones to potential base stations enables a timely AED
  delivery even in the unlikely situation of simultaneously having two SCA patients closely located to each other.
  Practitioners can benefit from this study in two respects. Firstly, we present
  mountain rescue services a novel approach for extending their traditional
  response system by an innovative means of transport. The technological features
  of the defibrillator drone itself can definitely increase the flexibility of
  response teams, thus reducing the time to provide critical support in the minutes
  after SCA. The developed model can provide a tool for decision makers in the
  field to optimally distribute defibrillator drones within local infrastructure
  under consideration of a given set of potential base stations. Notably, the proposed drone system
  should not replace air ambulance service, as the provision of intensive care
  medical treatment by helicopter crews is essential for the survival of the patient.
  From a scientific point of view, this paper is valuable in the sense that it
  is the first that studies the optimal allocation of defibrillator drones
  in mountainous regions. With this we enrich academic literature on the potential
  benefits, e.g. faster provision of emergency care to SCA patients, that arise
  from combining a transformative innovation and mathematical optimization.
  \par
  There are several issues that limit this research and that are worth further consideration.
  First of all, environmental factors that are characteristic for Alpine regions
  have great influence on the drones' performance. Huge differences in temperature
  and stormy conditions may have an impact on the operability of drones,
  e.g. less power supply by the battery if the air temperature is under
  \SI{-10}{\celsius}. In order to generate realistic data on the behavior of the considered drone model under such conditions, field experiments have to be conducted.
  Moreover, our study assumes that drones always fly in a straight line from
  the base station to the patient's site. From the analysis it became clear
  that high mountains account for long vertical take-off and landing times, which
  have a negative impact on the overall response times.
  A potential solution to this problem is to let the drone circle around
  high mountains instead of passing over them,
  which could result in reduced flight times of the drones.
  However, this may be subject to further research.
  \par
  This study solely focuses on minimizing the response time to deliver an AED to
  the patient's location by using a drone. Other times and factors that might
  contribute to a higher survival rate of the patient are disregarded in this setting.
  For instance, the time of the bystander's CPR initialization, the ability of the
  bystander to place the defibrillator paddles on the patient's chest,
  the quality of CPR or the connectivity of the bystander's mobile device
  to place an emergency call influence the survival rate.
  Other delays in time caused by the bystander's potential mental overload
  or panic could eventually have negative impacts on the patient's survival. It is to be noted this whole methodology only works if a bystander is on the scene.
  \par
  In our model, we assume that the flight corridor of the region is adapted to
  the parallel use of drones and ambulance helicopters.
  In practice, there is the need to equip drones and helicopters with automated
  collision warning systems to avoid aerial conflicts that might lead to the crash
  of both systems. GPS guidance of the drone, as assumed in our study, is also a
  limiting factor due to a lack of permission to fly beyond operator line of
  sight in almost all countries. Further political and legal discussions related
  to the automated use of drones for emergency purposes have to be stimulated. In this context, the Commission of the European Union has published a new regulation to harmonize remotely controlled UAV use in the entire European Union. This regulation was released in June 2019 and will become valid from July 2020 for all member states \cite{EUaircraft1, EUaircraft2,Huttunen2017}.
  Besides, in the current model it is obligatory for drones to land at the
  patient's site due to safety reasons to avoid harming the bystander or patient
  by moving rotor blades. In case the emergency location is covered by snow or
  trees or is situated in a canyon, a safe landing of the drone cannot be guaranteed.
  Here, parachute or drop free systems, i.e. using a winch to let an AED down on a rope,
  could help to supply AEDs in almost every situation.
  This model extension will also be considered in future work.
  \par
  In order to locate the drones at the defined base stations
  (i.e. shelter huts and fire rescue stations), infrastructural modifications
  must be implemented. Drones need electric power to operate.
  Consequently, shelter huts need to be equipped with power sources such as solar
  panels and accumulators that guarantee constant power supply to the drone.
  Autonomous take-off and landing systems also need to be installed in order
  to let the drone operate completely autonomously.
  Fire rescue stations need to be equipped with these systems too.
  Although we incorporate shelter huts and fire rescue stations to determine
  the optimal allocation of base stations, there may be other candidate locations
  that can be integrated in the model. Erecting new base stations at suitable locations could further reduce travel times to SCA patients. Prior to that, it has to be ensured that geographical conditions and legal regulations facilitate this. Experts (e.g. geologists, legal experts) have to be consulted, who decide which locations fulfill the desired requirements.
  Additionally, the installation of base stations in private homes and public institutions such as
  schools or post offices may be another option.
  Here, the number and type of base stations is mainly
  driven by financial considerations.
  \par
  Furthermore, we run each scenario with a homogeneous fleet of drones.
  Future research could address the combination of different drone types in order
  to better react to different requirements of the response situation, i.e. using
  cheaper and slower drones for patients that are located close to base stations and
  fast and expensive ones for far away patients. Other extensions could comprise more sophisticated flight routes that could potentially result in shorter flight times. However, this would require a full 3D model of the considered region, which is not available to us. Moreover, this approach would require identifying the shortest path through the 3D environment that is also safe in terms of external parameters (e.g. thermal up and down winds, wind in general, other obstacles such as ropeways). Another model extension could comprise a general budget constraint accounting for drone hosting costs (price of electricity, space etc.) to be included in the model. To properly estimate the costs, it needs further discussions with specialists in charge of maintaining the infrastructure on site and relevant technicians that have comprehensive knowledge on drone technology.
  Future work should also include
  a feasibility study. Bridging the gap between theory and practice by qualitative
  research is needed. This serves for identifying barriers and accelerators to
  turn the proposed concept of this study into practice.
  Hence, discussions with experts including mountain rescue services, EMS, and governmental
  stakeholders are necessary.
  Moreover, drone providers should also be consulted regarding the evaluation of the technological transferability of this approach.
  In conclusion, our model shows that drones might serve as a valuable compliment to already existing helicopter-based EMS due to theoretically achieved much faster response times.

\clearpage

  \begin{appendix}

    \section{Appendix} \label{sec:app}

    \begin{table}[ht]
      \footnotesize
      \centering
      \begin{tabular}{  l |r | r | r | r | r | r | r }
        \multicolumn{7}{l}{\textbf{Minimizing Number of Drones: }} \\
        \hline
        $t_{max}$ & s &  min($t$) & max($t$) & mean($t$) & median($t$) & 95 \% & 99 \% \\
        \hline \hline
        $14\colon 48$ & $18$ & $00\colon 31 $ & $14\colon 48 $ & $08\colon 11 $ & $ 08\colon 28 $ & $12\colon 59$ & $14\colon 13$ \\
        $15\colon 00$ & $17$ & $00\colon 47 $ & $14\colon 59 $ & $08\colon 01 $ & $ 08\colon 13 $ & $13\colon 13$ & $14\colon 43$  \\
        $15\colon 15$ & $16$ & $00\colon 38 $ & $15\colon 12 $ & $08\colon 18 $ & $ 08\colon 25 $ & $13\colon 41$ & $14\colon 44$ \\
        $15\colon 30$ & $15$ & $00\colon 31 $ & $15\colon 19 $ & $08\colon 39 $ & $ 08\colon 55 $ & $13\colon 57$ & $14\colon 53$ \\
        $15\colon 45$ & $15$ & $00\colon 43 $ & $15\colon 37 $ & $08\colon 41 $ & $ 08\colon 54 $ & $13\colon 59$ & $14\colon 53$ \\
        $16\colon 00$ & $14$ & $00\colon 49 $ & $15\colon 49 $ & $09\colon 04 $ & $ 09\colon 21 $ & $14\colon 15$ & $15\colon 19$ \\
        $16\colon 15$ & $14$ & $00\colon 31 $ & $16\colon 11 $ & $09\colon 15 $ & $ 09\colon 06 $ & $14\colon 42$ & $15\colon 43$ \\
        $16\colon 30$ & $12$ & $00\colon 33 $ & $16\colon 30 $ & $09\colon 29 $ & $ 09\colon 41 $ & $14\colon 34$ & $15\colon 43$ \\
        $16\colon 45$ & $12$ & $00\colon 31 $ & $16\colon 39 $ & $09\colon 21 $ & $ 09\colon 24 $ & $14\colon 49$ & $15\colon 56$ \\
        $17\colon 00$ & $11$ & $00\colon 31 $ & $16\colon 53 $ & $09\colon 51 $ & $ 10\colon 00 $ & $14\colon 54$ & $16\colon 06$ \\
        $17\colon 15$ & $11$ & $00\colon 42 $ & $17\colon 04 $ & $09\colon 46 $ & $ 09\colon 58 $ & $14\colon 53$ & $16\colon 06$ \\
        $17\colon 30$ & $11$ & $00\colon 33 $ & $17\colon 27 $ & $09\colon 38 $ & $ 09\colon 45 $ & $15\colon 12$ & $16\colon 20$ \\
        $17\colon 45$ & $11$ & $00\colon 33 $ & $17\colon 27 $ & $09\colon 59 $ & $ 10\colon 19 $ & $15\colon 17$ & $16\colon 24$ \\
        $18\colon 00$ & $11$ & $00\colon 33 $ & $17\colon 42 $ & $09\colon 56 $ & $ 10\colon 01 $ & $15\colon 37$ & $16\colon 52$ \\
        $18\colon 15$ & $11$ & $00\colon 31 $ & $18\colon 02 $ & $09\colon 42 $ & $ 09\colon 53 $ & $15\colon 18$ & $16\colon 54$ \\
        $18\colon 30$ & $10$ & $00\colon 59 $ & $18\colon 17 $ & $10\colon 04 $ & $ 10\colon 16 $ & $15\colon 38$ & $17\colon 18$ \\
        $18\colon 45$ & $11$ & $00\colon 59 $ & $18\colon 25 $ & $09\colon 28 $ & $ 09\colon 24 $ & $15\colon 21$ & $17\colon 33$ \\
        $19\colon 00$ & $10$ & $00\colon 51 $ & $18\colon 58 $ & $10\colon 11 $ & $ 10\colon 24 $ & $15\colon 58$ & $17\colon 57$ \\
        $19\colon 15$ & $10$ & $00\colon 56 $ & $19\colon 15 $ & $10\colon 29 $ & $ 10\colon 44 $ & $15\colon 53$ & $17\colon 23$ \\
        $19\colon 30$ & $10$ & $00\colon 33 $ & $19\colon 26 $ & $10\colon 18 $ & $ 10\colon 19 $ & $15\colon 58$ & $17\colon 44$ \\
        $19\colon 45$ & $10$ & $00\colon 31 $ & $19\colon 37 $ & $10\colon 24 $ & $ 10\colon 21 $ & $16\colon 24$ & $17\colon 58$ \\
        $20\colon 00$ & $10$ & $00\colon 31 $ & $19\colon 50 $ & $10\colon 33 $ & $ 10\colon 34 $ & $16\colon 30$ & $17\colon 55$ \\
        \hline

      \end{tabular}
      \caption{Table summarizing the generated results when minimizing the number of defibrillator drones using \textit{LifeDrone AED} system  ($\text{mm}\colon \text{ss}$).}
      \label{tab:minDrones}
    \end{table}

    \begin{table}[ht]
      \footnotesize
      \centering
      \begin{tabular}{  l |r | r | r | r | r | r | r }

        \multicolumn{7}{l}{\textbf{Minimizing Number of Drones:}} \\
        \hline
        $t_{max}$ & s &  min($t$) & max($t$) & mean($t$) & median($t$) & 95 \% & 99 \% \\
        \hline \hline
        $06\colon 41$ & $19$ & $00\colon 20 $ & $06\colon 41 $ & $03\colon 47 $ & $ 03\colon 52 $ & $05\colon 58$ & $06\colon 34$  \\
        $06\colon 45$ & $18$ & $00\colon 20 $ & $06\colon 45 $ & $03\colon 48 $ & $ 03\colon 51 $ & $06\colon 07$ & $06\colon 36$ \\
        $07\colon 00$ & $16$ & $00\colon 20 $ & $06\colon 56 $ & $03\colon 56 $ & $ 03\colon 56 $ & $06\colon 26$ & $06\colon 49$ \\
        $07\colon 15$ & $14$ & $00\colon 20 $ & $07\colon 15 $ & $04\colon 10 $ & $ 04\colon 19 $ & $06\colon 28$ & $06\colon 52$ \\
        $07\colon 30$ & $14$ & $00\colon 22 $ & $07\colon 23 $ & $04\colon 08 $ & $ 04\colon 11 $ & $06\colon 33$ & $07\colon 05$ \\
        $07\colon 45$ & $12$ & $00\colon 20 $ & $07\colon 44 $ & $04\colon 26 $ & $ 04\colon 28 $ & $06\colon 48$ & $07\colon 23$ \\
        $08\colon 00$ & $11$ & $00\colon 20 $ & $07\colon 46 $ & $04\colon 32 $ & $ 04\colon 34 $ & $06\colon 55$ & $07\colon 25$ \\
        $08\colon 15$ & $11$ & $00\colon 26 $ & $08\colon 14 $ & $04\colon 35 $ & $ 04\colon 39 $ & $07\colon 06$ & $07\colon 44$ \\
        $08\colon 30$ & $10$ & $00\colon 22 $ & $08\colon 29 $ & $04\colon 48 $ & $ 04\colon 52 $ & $07\colon 23$ & $07\colon 59$ \\
        $08\colon 45$ & $11$ & $00\colon 26 $ & $08\colon 36 $ & $04\colon 38 $ & $ 04\colon 42 $ & $07\colon 12$ & $07\colon 51$ \\
        $09\colon 00$ & $10$ & $00\colon 22 $ & $08\colon 49 $ & $04\colon 47 $ & $ 04\colon 44 $ & $07\colon 25$ & $08\colon 16$ \\
        $09\colon 15$ & $8$ & $00\colon 32 $ & $09\colon 15 $ & $05\colon 23 $ & $ 05\colon 31 $ & $08\colon 10$ & $08\colon 53$ \\
        $09\colon 30$ & $8$ & $00\colon 24 $ & $09\colon 28 $ & $05\colon 13 $ & $ 05\colon 16 $ & $08\colon 27$ & $09\colon 03$ \\
        $09\colon 45$ & $7$ & $00\colon 26 $ & $09\colon 45 $ & $05\colon 36 $ & $ 05\colon 46 $ & $08\colon 42$ & $09\colon 21$ \\
        $10\colon 00$ & $7$ & $00\colon 26 $ & $09\colon 43 $ & $05\colon 28 $ & $ 05\colon 34 $ & $08\colon 40$ & $09\colon 22$ \\
        $14\colon 48$ & $3$ & $00\colon 32 $ & $13\colon 52 $ & $08\colon 14 $ & $ 08\colon 22 $ & $12\colon 00$ & $13\colon 20$ \\
        $15\colon 00$ & $3$ & $00\colon 32 $ & $14\colon 57 $ & $08\colon 24 $ & $ 08\colon 49 $ & $12\colon 30$ & $13\colon 46$ \\
        \hline

      \end{tabular}
      \caption{Table summarizing the generated results when minimizing the number of defibrillator drones using
      the \textit{Wingcopter 178} system   ($\text{mm}\colon \text{ss}$).}
      \label{tab:minDronesfast}
    \end{table}

    \begin{figure}
      \begin{center}
        \includegraphics[width=\textwidth]{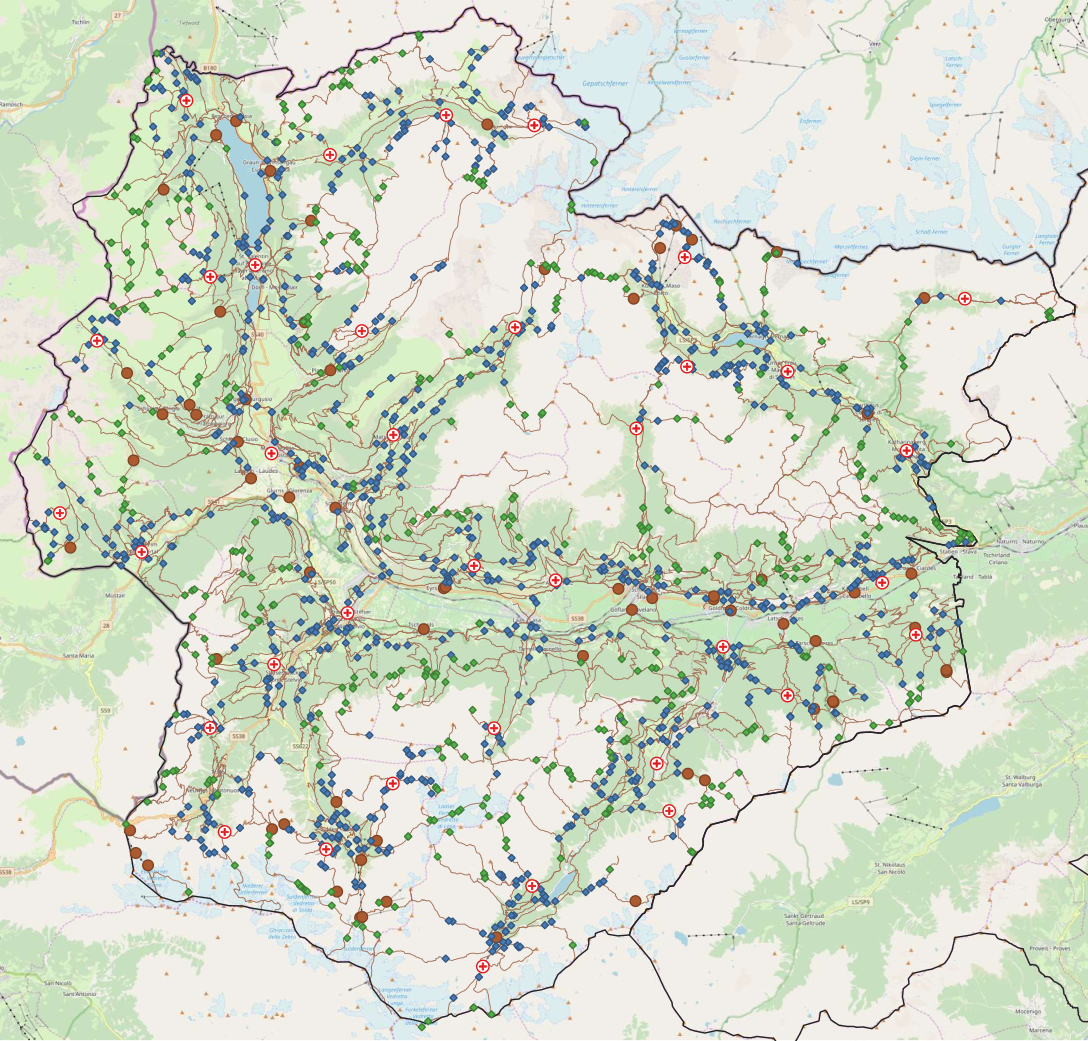}
        \caption{Illustration of the solution of configuration \texttt{ATTs36} using
        the \textit{LifeDrone AED}.}
        \label{fig:mapd36slow}
      \end{center}
    \end{figure}

    \begin{figure}
      \begin{center}
        \includegraphics[width=\textwidth]{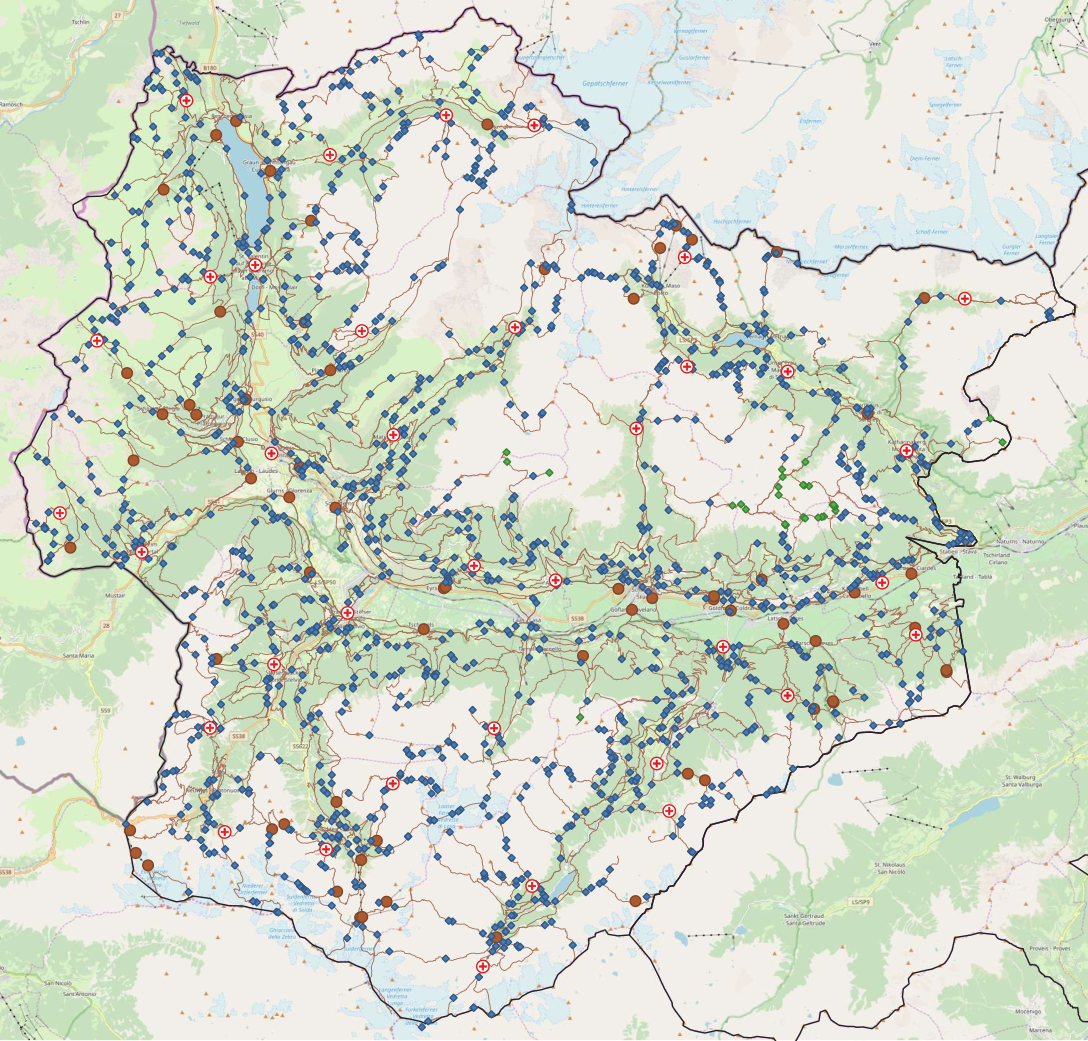}
        \caption{Illustration of the solution of configuration \texttt{ATTs36} using
        the \textit{Wingcopter 178} system. }
        \label{fig:mapd36fast}
      \end{center}
    \end{figure}

    \begin{figure}
      \begin{center}
        \includegraphics[width=\textwidth]{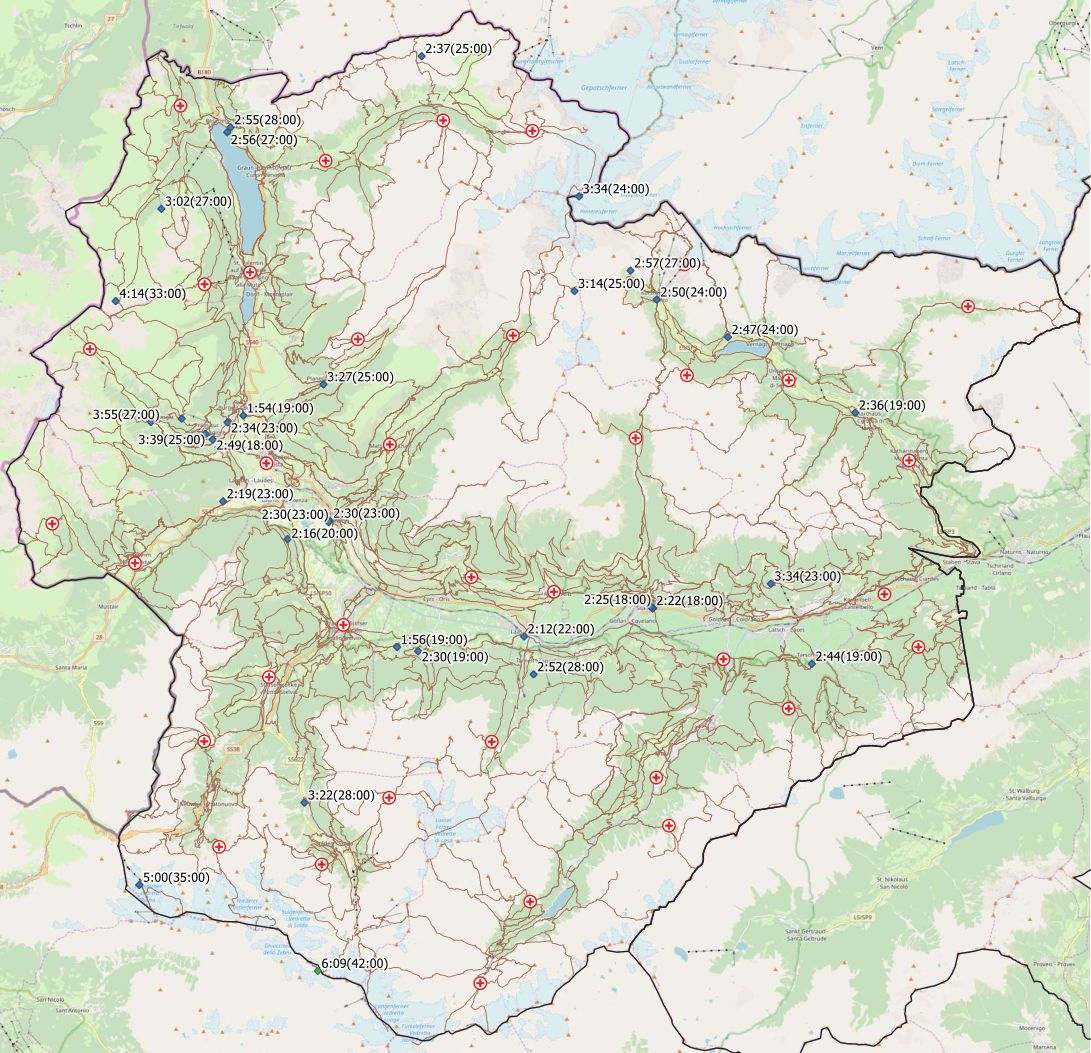}
        \caption{The figure shows selected patients' locations from the helicopter data set.
        Additionally, we report the flight times  ($\text{mm}\colon \text{ss}$) of the defibrillator drones and the
        historical helicopter flight times (in brackets).}
        \label{fig:maphelidrone}
      \end{center}
    \end{figure}

  \end{appendix}

  \end{document}